\documentclass[]{aa} 

\graphicspath{./images/}  
\usepackage{graphicx}
\usepackage{txfonts}
\usepackage{environ}
\usepackage{booktabs}
\usepackage{floatrow}
\usepackage{gensymb}
\usepackage{multirow}
\usepackage{float}
\floatstyle{plaintop}
\restylefloat{table}
\usepackage[colorlinks=true,linkcolor=blue,citecolor=blue]{hyperref}

\def\ergs{{erg~s$^{-1}$}}

\def\msun{{$M_\odot$}}

\def\lbol{{$L_{\rm Bol}$}}

\def\pds{PDS~456}

\def\co32{{CO(3$-$2)}}
\def\kms{{km~s$^{-1}$}}

\def\mout{{$M_{\rm mol}^{\rm out}$}}
\def\mdot{{$\dot{M}_{\rm of}$}}
\def\pdot{{$\dot{P}_{\rm of}$}}
\def\mdotm{{$\dot{M}_{\rm mol}$}}
\def\pdotm{{$\dot{P}_{\rm mol}$}}
\def\msunyr{{$M_{\odot}$~yr$^{-1}$}}

\begin{document}


\title{The gentle monster PDS 456:}
\subtitle{the kpc-scale molecular outflow and its implications for QSO feedback}


\author{M. Bischetti \inst{1}
         \and E. Piconcelli \inst{1}
         \and C. Feruglio \inst{2}
         \and F. Fiore \inst{2}
         \and S. Carniani \inst{3}
         \and M. Brusa \inst{4,5}
        \and C. Cicone \inst{6}
        \and C. Vignali \inst{4,5}
        \and A. Bongiorno \inst{1}
        \and G. Cresci \inst{7}
        \and V. Mainieri \inst{8}
        \and R. Maiolino \inst {9,10}
        \and A. Marconi \inst{7,11}
        \and E. Nardini \inst{7}
        \and L. Zappacosta \inst{1}
        }

 \titlerunning{ALMA observation of PDS 456}\authorrunning{M.~Bischetti et al.}
 
\institute{INAF - Osservatorio Astronomico di Roma, Via Frascati 33, I--00078 Monte Porzio Catone (Roma), Italy
        \and INAF - Osservatorio Astronomico di Trieste, via G.B. Tiepolo 11, I--34143 Trieste, Italy
        \and Scuola Normale Superiore, Piazza dei Cavalieri 7, I--56126, Pisa (PI), Italy
        \and Dipartimento di Fisica e Astronomia, Universit\`a di Bologna, via Gobetti 93/2, I--40129 Bologna, Italy
        \and INAF - Osservatorio di Astrofisica e Scienza dello Spazio di Bologna, via Gobetti 93/3, I--40129 Bologna, Italy
        \and INAF - Osservatorio Astronomico di Brera, Via Brera 28, I--20121 Milano, Italy
        \and INAF - Osservatorio Astrofisico di Arcetri, Largo E. Fermi 5, I--50125, Firenze, Italy
        \and European Southern Observatory, Karl-Schwarzschild-str. 2, 85748 Garching bei M\"{u}nchen, Germany
        \and Cavendish Laboratory, University of Cambridge, 19 J. J. Thomson Avenue, Cambridge CB3 0HE, UK
   	    \and Kavli Institute for Cosmology, University of Cambridge, Madingley Road, Cambridge CB3 0HA, UK
   	    \and Dipartimento di Fisica e Astronomia, Universit\`a di Firenze, Via G. Sansone 1, I--50019, Sesto Fiorentino (Firenze), Italy
        }

\abstract{We report on the first ALMA observation of the \co32\ and rest-frame $\sim340$ GHz continuum emission in \pds, which is the most luminous, radio-quiet QSO in the local Universe ($z\simeq0.18$), with a bolometric luminosity \lbol$\sim10^{47}$ \ergs. ALMA angular resolution allowed us to map scales as small as $\sim700$ pc.
The molecular gas reservoir, traced by the core of the very bright \co32\ emission line, is distributed in a compact rotating disk, with size of $\sim1.3$ kpc, seen close to face-on ($i\sim25$ deg).  Fast  \co32\ emission in the velocity range $v\in[-1000,500]$ \kms\ is also present. Specifically, we detect several blue-shifted clumps out to $\sim5$ kpc from the nucleus, in addition to a compact (R$\lesssim1.2$ kpc), broad emission component.  These components reveal a galaxy-wide molecular outflow, with a total mass \mout$\sim2.5\times10^8$ \msun\ (for an $\alpha_{\rm CO}=0.8$ \msun(K \kms\ pc$^2$)$^{-1}$) and a mass outflow rate \mdotm$\sim290$ \msunyr. The corresponding depletion time is $\tau_{\rm dep}\sim8$ Myr, shorter than the rate at which the molecular gas is converted into stars, indicating that the detected outflow is potentially able to quench star-formation in the host.
The momentum flux of the molecular outflow normalised to the radiative momentum output (i.e. $L_{\rm Bol}/c$)  is $\lesssim$ 1, comparable to that of the X-ray ultra-fast outflow (UFO) detected in \pds. This is at odds with the expectations for an energy-conserving expansion suggested for most of the large-scale outflows detected in low-luminosity AGN so far. 
We suggest three possible scenarios that may explain this observation: (i) in very luminous AGN such as our target the molecular gas phase is tracing only a fraction of the total outflowing mass; (ii) a small coupling between the shocked gas by the UFO and the host-galaxy ISM (iii) AGN radiation pressure may play an important role in driving the outflow.}

\keywords{}

\maketitle


\section{Introduction}

The coupling between the super-massive black hole (SMBH) energy output and the interstellar and circum-galactic medium (ISM and CGM) of the host-galaxy is  still an open issue,  particularly relevant for  hyper-luminous quasi-stellar objects (QSOs) with SMBH mass  $\geq$ 10$^{9}$  M$_\odot$ and bolometric luminosity $L_{\rm Bol}> 10^{47}$ \ergs, i.e. at the  brightest end of the active galactic nuclei (AGN) luminosity function. 
Mechanical and radiative QSO-driven feedback processes have been implemented in models of galaxy evolution to prevent massive galaxies from over-growing, change their colours,  heat both ISM and CGM and enrich them with metals \citep[e.g.][and references therein]{Croton06, Sijacki07, Voit15, Gaspari17, Choi18}. 

The impressive growth in the number of QSO-driven outflows discovered in the last decade represents a great advance in our comprehension of the feedback process. These outflows have been detected in all gas phases and at all spatial scales \citep[sub-pc to several kpc, see][and references therein]{Fiore17}, and provide  a very promising mechanism to efficiently deposit energy and momentum into the surrounding gas \citep[e.g.][]{Faucher-Giguere12, Zubovas&King12}, with the most powerful ones exhibiting a kinetic power up to a few percent of \lbol\ \citep[e.g.][]{Feruglio10,Maiolino12,Cicone14,Aalto15,Bischetti17}. 
In several AGN, mainly in the cold molecular and neutral gas phases,  mass outflow rates fairly exceeding the star formation rate have been measured \citep[e.g.][]{Feruglio13b, Alatalo15a, Alatalo15, Cicone15, Fluetsch19}, indicating that these outflows may affect the evolution of the host galaxy.

Ultra-fast outflows (UFOs) of highly ionised gas observed at sub-pc scales \citep{Reeves03,Tombesi12} have been proposed as the likely origin of galaxy-wide outflows, interpreted as the result of the impact of UFOs on the ISM \citep[][and references therein]{King&Pounds15}
Furthermore,
both models and observations of kpc-scale outflows seem to indicate a  UFO-ISM interaction in an energy-conserving regime, whereby the swept-up gas expands adiabatically.
So far the co-existence of a massive molecular outflow with a nuclear UFO has been confirmed in a handful of AGN with \lbol$\sim10^{44}- 10^{46}$ \ergs\ \citep{Tombesi15,Feruglio15,Longinotti15} and in APM 08279$+$5255 \citep{Feruglio17}, which is
a gravitationally-lensed QSO at $z$ $\sim$ 4 with an estimated intrinsic \lbol\ of few times  10$^{47}$ \ergs\
\citep{Saturni18}.
In all these sources the momentum boost (i.e. the momentum flux of the wind normalised to the AGN radiative momentum output, \lbol/$c$)
of the UFO is $\sim$ 1, while the momentum rate of the molecular outflow is usually $\gg$ 1, in qualitative agreement with the theoretical predictions for an energy-conserving expansion \citep{Faucher-Giguere12,Costa14}. However, these results are still limited to a very small sample and suffer from large observational uncertainties, mostly due to the relatively low signal-to-noise of the UFO- or outflow-related features confirmed in spectra, or to the limited spatial resolution of sub-mm observations. Recent works increasing the statistics of sources with detection of molecular outflows have widened the range of measured energetics \citep[e.g.][]{GarciaBurillo14,Veilleux17,Feruglio17,Brusa18,BarcosMunoz18,Fluetsch19}, consistently with 
driving mechanisms alternative to the energy-conserving expansion, such as direct radiation pressure onto the host-galaxy ISM \citep[e.g.][]{Ishibashi&Fabian14,Ishibashi18,Costa18a}.


In order to study the interplay between UFOs and large-scale outflows in the still little explored high-\lbol\ regime, we have targeted with ALMA the QSO \pds, which is the most luminous, radio-quiet AGN (\lbol\ $\sim10^{47}$ \ergs) in the local Universe at $z\simeq0.18$ \citep{Torres97,Simpson99}. This has allowed us to probe the molecular gas reservoir in a hyper-luminous QSO with unprecedented spatial resolution ($\sim700$ pc).
\pds\ exhibits the prototype of massive and persistent UFO detected in the X-rays, identified as a quasi-spherical wind expanding with a velocity of $\sim0.3c$ and kinetic power of $\sim20-30$\% of $L_{\rm Bol}$ \citep{Nardini15, Luminari18}, arising at $\sim$ 0.01 pc from the SMBH.
\citet{Reeves16} have reported the discovery of a complex of soft X-ray broad absorption lines, possibly associated with
a lower ionisation, decelerating ($\sim$ 0.1$c$) phase of the UFO out to pc scales.
Moreover, \citet{Hamann18} have recently claimed the presence of highly blueshifted CIV absorption line in the {\it Hubble Space Telescope} UV spectra of \pds, tracing an  outflow with velocity of 0.3$c$,  similar to that measured for the UFO. 

Given its uniqueness in terms of presence of very fast outflows observed in several wavebands and its high luminosity, which makes it a local counterpart of the hyper-luminous QSOs shining at $z\sim2-3$, \pds\  stands out as one of the best targets to investigate the presence of a molecular outflow and the effects of the QSO activity on the host-galaxy ISM.
Nonetheless, the properties of the molecular gas of \pds\  have been poorly studied so far,  being based on  a  low-resolution ($7\times4.8$ arcsec$^2$) and low-sensitivity observation performed with the OVRO array \citep[][hereafter Y04]{Yun04}.
The detection of a CO(1-0) emission line with a FWHM = 180 \kms\ and line flux of  $\sim$ 1.5  Jy \kms\  implies a molecular gas reservoir of few times $10^{9}$ \msun, which 
 is an intermediate value between those typically measured for
  {\it blue} Palomar-Green QSOs and local ultra-luminous infrared galaxies (ULIRGs) \citep[e.g.][]{Solomon97,Evans06,Xia12}. 
The $K$-band image obtained at the \textit{Keck} Telescope shows three compact sources detected at $\sim3$ arcsec from the QSO, suggesting the possible presence of  companions at a projected distance of $\sim9$ kpc (Y04).
 
The paper is organised as follows. In Sect. 2 we describe the ALMA observation of \pds\ and the data reduction procedure. Our analysis and results are presented in  Sect. 3.
We discuss in Sect. 4 and conclude our findings in Sect. 5.
At the redshift of \pds, the physical scale is $\sim3.1$ kpc arcsec$^{-1}$, given a $H_0=69.6$, $\Omega_{\rm m}=0.286$, and $\Omega_{\rm \Lambda}=0.714$ cosmology.



\begin{figure*}[thb]
    \centering
    \includegraphics[width = 1\textwidth]{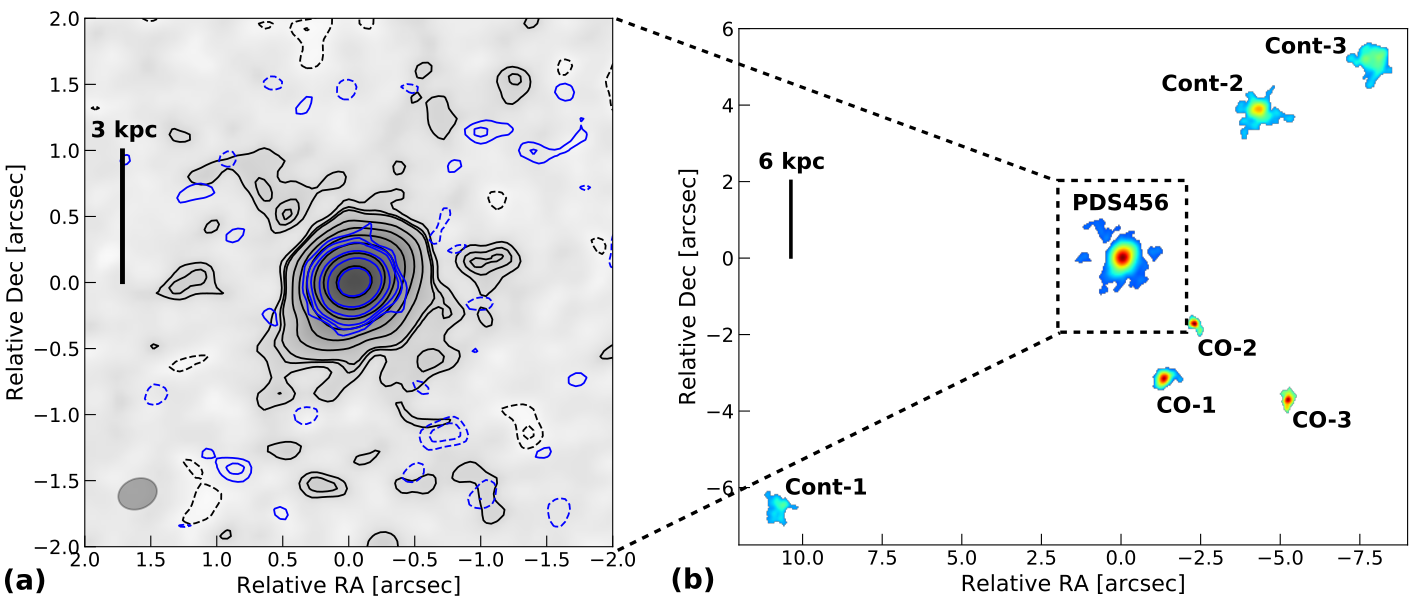}
    \caption{Panel \textbf{\textit{(a):}} map of the continuum-subtracted, \co32\ emission of \pds\, integrated over a linewidth of 320 \kms. Black contours indicate the [-3,-2, 2, 3, 2$^n$]$\sigma$ significance levels ($n\geq2$ and $\sigma=0.013$ Jy beam$^{-1}$ \kms) of the \co32\ emission.
    Blue contours indicate the (rest-frame) $\sim340$ GHz continuum  [-3,-2, 2, 3, 2$^n$]$\sigma$ levels (with $\sigma=9.6$ $\mu$Jy beam$^{-1}$). The ALMA synthetic beam is indicated by the grey ellipse. Panel \textbf{\textit{(b):}} map of the line and continuum emitters detected in the ALMA field of view at $\gtrsim5\sigma$.
    }
    \label{fig:CO-cont-map}
\end{figure*}

\section{ALMA observation and data reduction}\label{sect:datared}
We present in this work the ALMA Cycle 4 observation (project 2016.1.01156.S, P.I. E. Piconcelli) of \pds, performed on 5 May 2017 for a 4.1 hours on-source integration time. The ALMA array was arranged in C40-5 configuration, with a maximum projected baseline of $\sim1.1$ km. We used the ALMA band 7 receiver and the frequency division mode of the ALMA correlator, providing us with four spectral windows of 1.875 GHz width and a spectral resolution of 31.25 MHz ($\sim30$ \kms). One spectral window (spw0) was centred at 292 GHz to cover the expected frequency of the \co32\ emission  (rest frequency 345.796 GHz), based on the [Fe II] redshift $z_{\rm [FeII]}=0.184$ from \citet{Simpson99}. The second spectral window (spw1) was set adjacent to the first with $\sim300$ MHz overlap on the lower frequency side to accurately estimate the continuum. The sideband not including \co32\ emission with the two remaining spectral windows was set at $\sim280$ GHz.
Visibilities were calibrated by using the CASA 4.7.2 software \citep{McMullin07} in the pipeline mode and the default calibrators provided by the Observatory: bandpass calibrator J175132$+$093958 (band 7 flux 1.42$\pm$0.07 Jy), flux and phase calibrator J173302$-$130445 (band 7 flux 1.12$\pm$0.06 Jy), water vapour radiometer calibrator J173811$-$150300 (band 3 flux 0.11$\pm$0.01 Jy). The absolute flux accuracy is better than 10\%.

To estimate the rest-frame $\sim340$ GHz continuum emission we averaged the  visibilities in the four spectral windows excluding the spectral range covered by the \co32\ emission ($\sim1$ GHz). 
Moreover, to accurately model the continuum emission close to the \co32\ line, we performed a combined fit of only spw0 and spw1 in the UV plane. We did not include the lower sideband to avoid introducing systematics usually associated with the relative calibration of distant spectral windows. 
The relative flux calibration of spw0 and spw1 was verified for all calibrators and, for \pds, in the overlap range of the two spectral windows.
The agreement of the continuum levels in the overlap region is better than 2\%.
As the intrinsic QSO continuum variation across spw0 and spw1 is expected to be less than 1\%, we fitted a zero order model in the UV plane to the continuum channels ($|v|>1000$ \kms\ from the peak of the \co32 emission line).  A first order polynomial fit to the continuum emission did not significantly change our results. We subtracted this fit from the global visibilities and created continuum-subtracted \co32\ visibilities.

We investigated different cleaning procedures to produce the continuum-subtracted cube of \co32. We preferred the Hogbom algorithm and the application of interactive cleaning masks for each channel and cleaning iteration. The usage of the Clark cleaning algorithm does not significantly affect the properties of the QSO emission but increases the number of negative artefacts in the ALMA maps, while non-interactive cleaning (without masks) results in positive residuals at the location of the QSO. We chose a natural weighting, a cleaning threshold equal to the rms per channel, a pixel size of 0.04 arcsec and a spectral resolution of $\sim30$ \kms. The final beam size of the observation is ($0.23\times0.29$) arcsec$^2$ at position angle $\rm PA= -70$ deg. The 1$\sigma$ rms in the final cube is $\sim 0.083$ mJy beam$^{-1}$ for a channel width of 30 \kms.
By adopting the same deconvolution procedure as explained above, we obtain a continuum map with synthetic beam of ($0.24\times0.30$) arcsec$^2$ and rms of 9.6 $\mu$Jy beam$^{-1}$ in the aggregated bandwidth.
We also produced a \co32\ data-cube with increased angular resolution by applying a briggs weighting to the visibilites in our ALMA observation with robust parameter $b=-0.5$, resulting into an ALMA beamsize of $0.16\times0.19$ arcsec$^2$ and a rms sensitivity of $\sim 0.16$ mJy beam$^{-1}$ for a 30 \kms\ channel width.

\section{Results}\label{sect:results}

\begin{figure}[htb]
    \caption{Continuum-subtracted spectrum of the \co32\ emission line in \pds, extracted from a circular region of 1 arcsec radius. Panel \textbf{\textit{(a)}} shows the integrated flux density as a function of velocity, corresponding to a circular region of 1 arcsec radius centred on the QSO position. The channel width is 30 \kms. The inset \textbf{\textit{(b)}} shows a zoomed-in view of the high-velocity wings in the \co32\ line profile. }
    \label{fig:spectrum}
    \includegraphics[width=1\columnwidth]{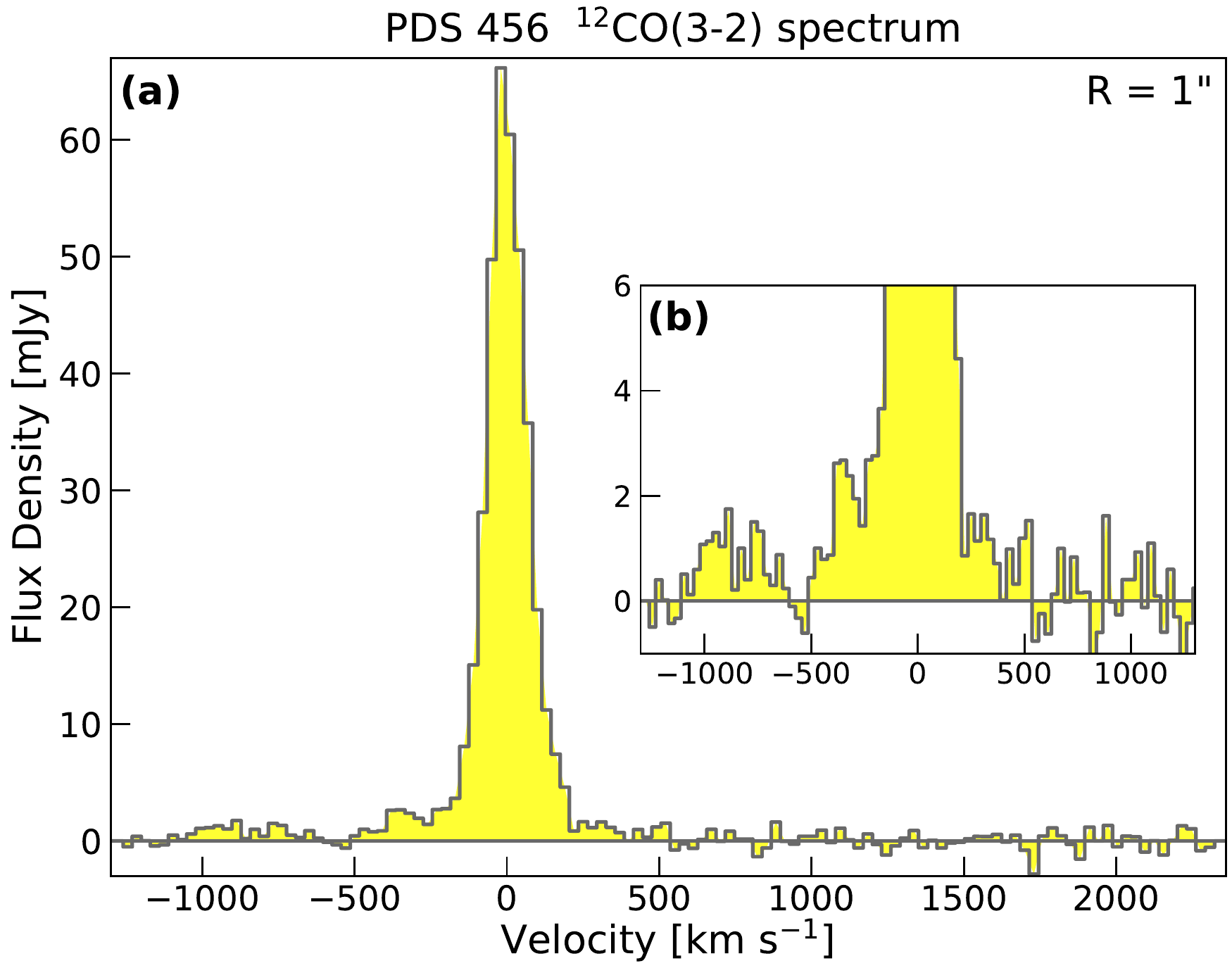}

\end{figure}

The continuum-subtracted, velocity integrated \co32\ emission in the velocity range $v \in$ [-160,160] \kms\ of \pds\ is shown in Fig. \ref{fig:CO-cont-map}a. The contours of the 340 GHz continuum emission are also plotted. 
The peak of the \co32\ emission is located at  RA (17:28:19.79 $\pm$ 0.01), Dec (-14:15:55.86 $\pm$ 0.01), consistent with the position of \pds\  based on  VLA data (Y04). The \co32\ emission line is detected with a significance of $\sim350\sigma$ (with  $\sigma= 0.013$ Jy beam$^{-1}$ \kms). The bulk of the emission is located within $\sim1$ arcsec from the QSO, with some extended, fainter clumps located at larger distance detected with statistical significance of $\sim4\sigma$. 
Both \co32\ and continuum emission are resolved by the ALMA beam in our observation. Specifically, a two-dimensional Gaussian fit of \co32\ in the image plane results into a deconvolved FWHM size of (0.28$\pm$0.02) $\times$ (0.25$\pm$0.02) arcsec$^2$, which corresponds to a physical size of $\sim0.9$ kpc. 
A fit of the continuum map gives a FWHM deconvolved size of (0.19$\pm$0.02) $\times$ (0.17$\pm$0.02) arcsec$^2$ and a flux density of $0.69\pm0.02$ mJy.  

In addition to \pds, three line emitters (CO-1, CO-2, CO-3) and three continuum emitters (Cont-1, Cont-2, Cont-3) are detected at $\gtrsim$5$\sigma$ in the ALMA primary beam ($\sim20$ arcsec), as displayed in Fig. \ref{fig:CO-cont-map}b. The proximity in sky frequency of the line emitters suggests that these are \co32\ emitters located at approximately the same redshift of the QSO. A detailed analysis of the galaxy overdensity around \pds\ will be presented in a forthcoming paper (Piconcelli et al. 2019 in prep.).

\begin{figure*}[htb]
\centering
\includegraphics[width = 0.515\textwidth]{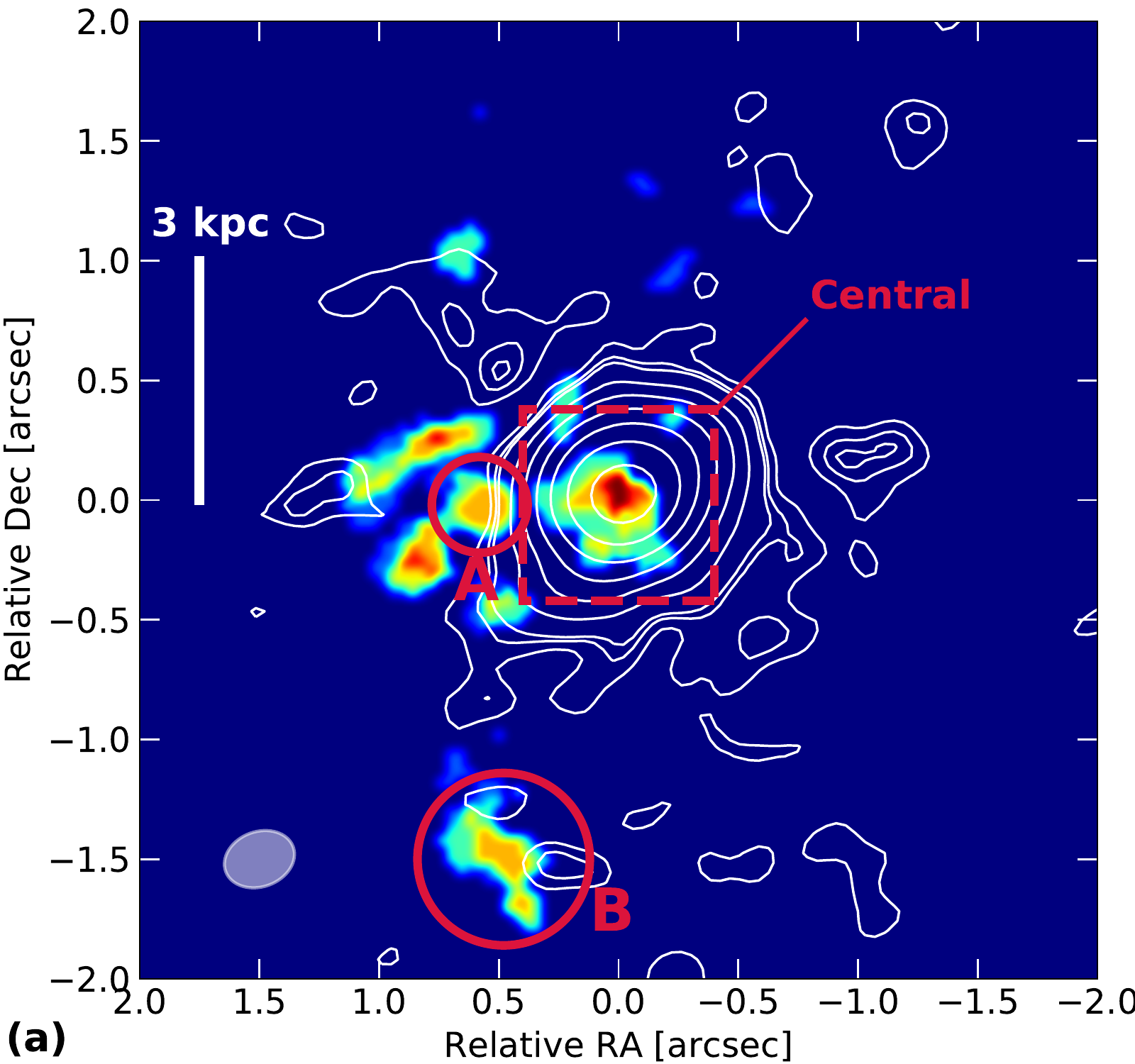}
\includegraphics[width = 0.475\textwidth]{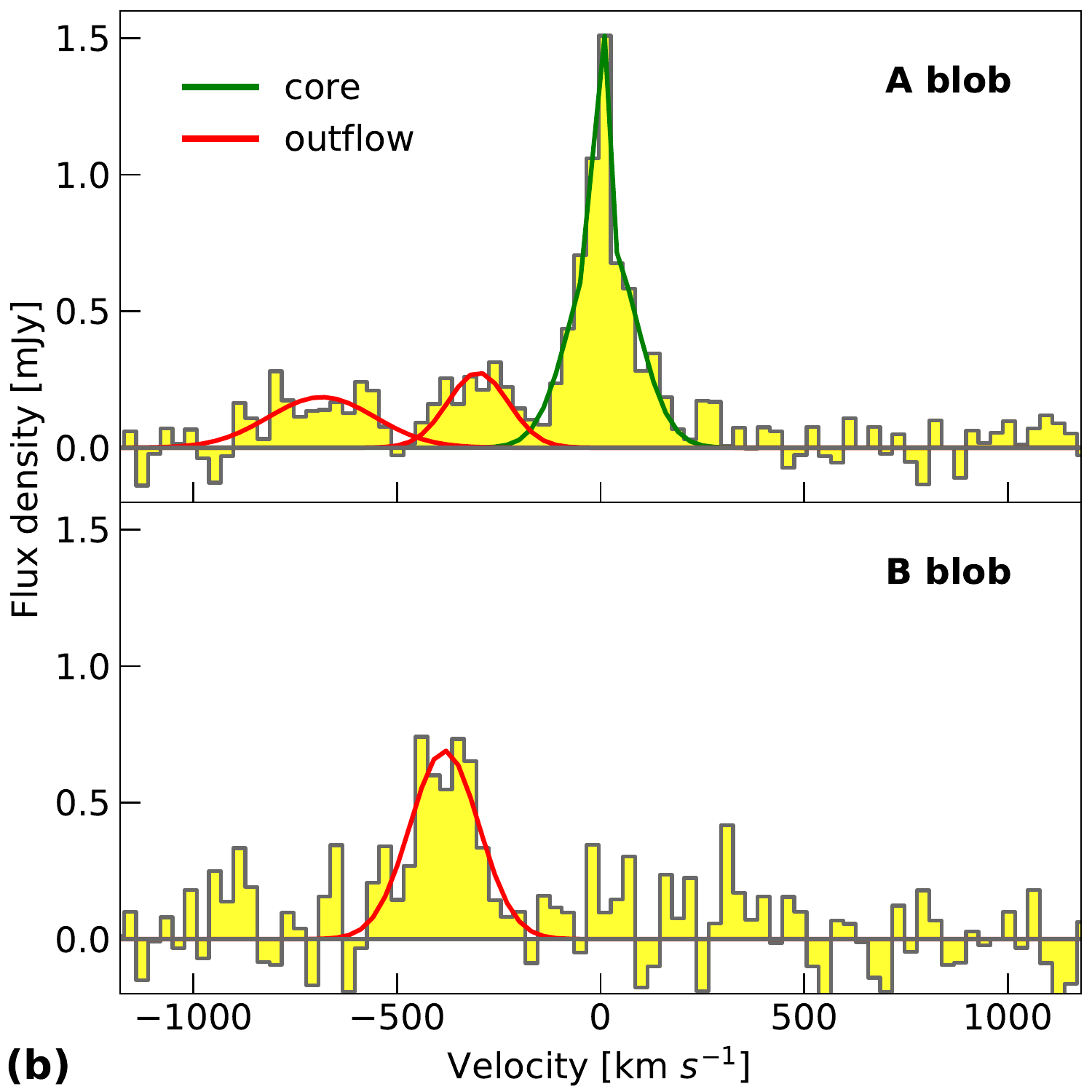}
\caption{Panel \textit{\textbf{(a)}}: velocity integrated map of the blue-shifted  ($v<-250$ \kms) \co32\ emission obtained by integrating the emission detected at $>3\sigma$ in each 30 \kms\ spectral channel, for at least four contiguous  channels (i.e. over a velocity range of $\geq$ 120 \kms). White contours show the systemic \co32\ emission (same as Fig. 1a). Panel \textit{\textbf{(b)}}: \co32\ spectra of the extended outflowing clumps A and B shown in panel (a), together with their best-fit multi-Gaussian components model. The spectrum extracted at the position of clump A (top panel), located at $\sim1.8$ kpc from the QSO, shows \co32\ emission centred at $v\sim0$ (systemic emission from the QSO, green curve) plus two components with blueshifted velocities of $v\sim-300$ and $\sim-700$ \kms. 
The spectrum extracted at the position of clump B (bottom panel) shows no contamination from the \co32\ systemic emission, while blueshifted emission is detected at $v\sim-400$ \kms. }
\label{fig:mappa-broad}
\end{figure*}

\begin{figure}[htb]
    \centering
    \includegraphics[width=1\columnwidth]{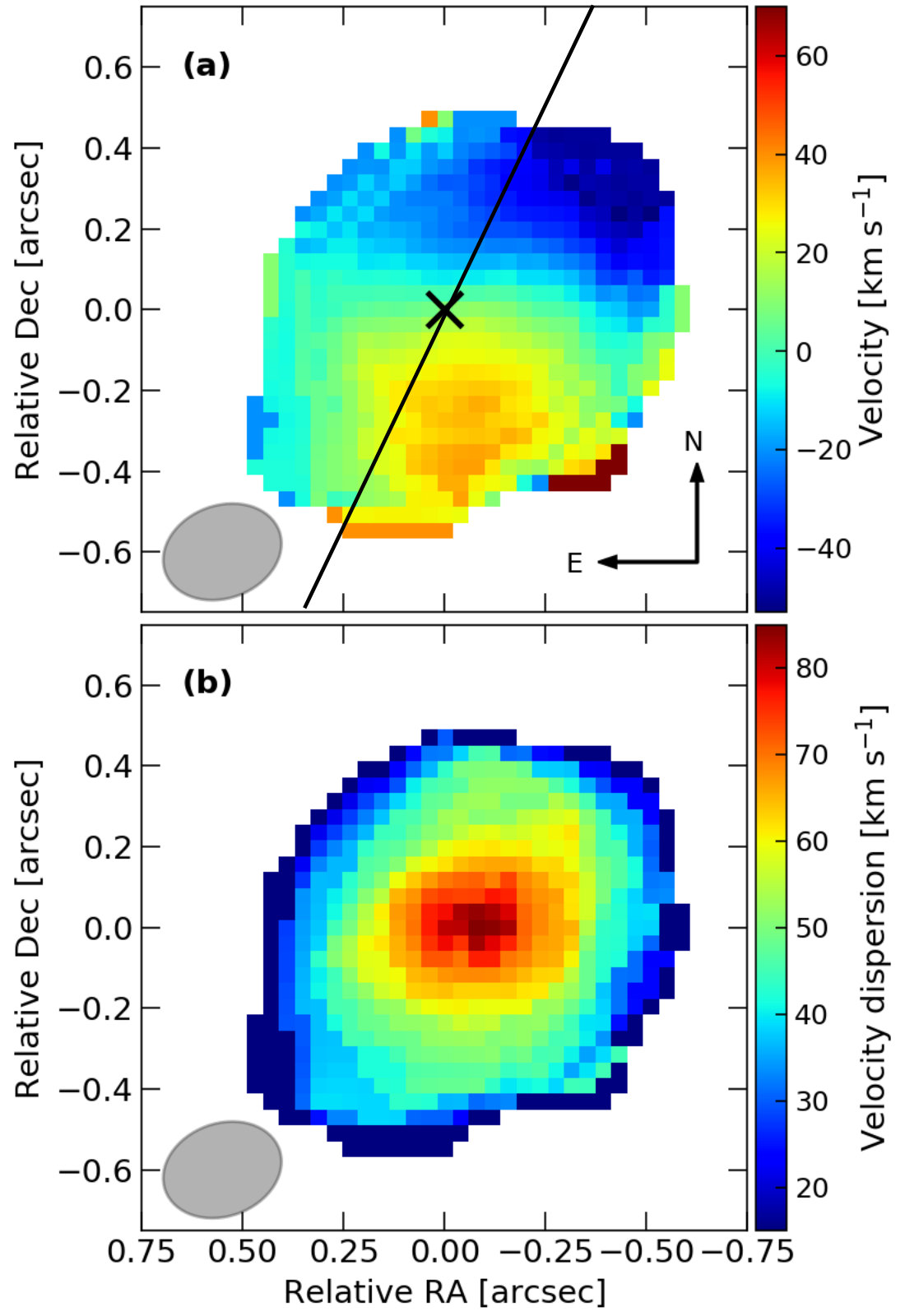}
    \caption{Panel \textit{\textbf{(a)}} shows the velocity map of the \co32 emission detected at $\gtrsim5\sigma$ in the host galaxy of \pds, resolved with $\sim12$ ALMA beams (indicated by the grey ellipse).  The main kinematic axis is in indicated by the black line. Panel \textit{\textbf{(b)}} displays the velocity dispersion map, characterised by values of $\sigma\lesssim80$ \kms.}
    \label{fig:core-moments}
\end{figure}

As detailed in Sect. \ref{sect:datared}, for an accurate study of the \co32\ emission line profile and estimation of the underlying continuum, we combine two adjacent spectral windows in order to exploit the largest possible spectral coverage (i.e. $\sim$ 3.8 GHz).
In Fig. \ref{fig:spectrum}  we present the continuum-subtracted spectrum of the \co32\ emission in \pds, extracted from a circular region of 1 arcsec radius. By fitting the line with a single Gaussian component, we measure a peak flux density S$_{\rm 3-2} = 63.6 \pm 0.7$ mJy and a FWHM $= 160\pm30$ \kms. The line peak
corresponds to a redshift $z_{\rm CO} = 0.1850\pm0.0001$,  consistent with the CO(1$-$0) based redshift from Y04, but significantly larger than the value $z_{\rm [FeII]} = 0.1837\pm0.0003$ derived from the [FeII] emission line in the near-IR spectrum \citep{Simpson99}. 
We find a line brightness ratio S$_{\rm 3-2}/$S$_{\rm 1-0}\sim8$ (computed by using the $S_{\rm 1-0}$ flux density reported by Y04\footnote{We estimate a possible contamination to the CO(1$-$0) flux due to the companion sources CO-1 and CO-2 to be $\lesssim2$\%, once a CO excitation ladder similar to the Galactic one is assumed \citep{Carilli13}.}) in agreement with the CO excitation ladder typically found in QSOs \citep{Carilli13}. This suggests that our ALMA observation is able to recover the bulk of the \co32\ emission in \pds. According to our observation setup, the largest recoverable scale is $2.2$ arcsec ($\sim$6.6 kpc), fairly larger than the size of the CO emission measured in local luminous infrared galaxies and QSOs \citep[e.g.][]{Bryant&Scoville99,Solomon05,Moser16}.
We derive an integrated intensity S$\Delta$v$_{\rm 3-2} = 10.6 \pm 0.2$ Jy \kms. This translates into a luminosity L$^\prime$CO$_{3-2}= 2.1\times10^9$ K \kms\ pc$^2$ and a luminosity ratio L$^\prime$CO$_{3-2}/$L$^\prime$CO$_{1-0}\sim 0.85$.

The line profile of the \co32\ emission exhibits a blue tail,  indicating the presence of emitting material with velocities down to $\sim$ $-$1000 \kms\ (see Fig. \ref{fig:spectrum}b), that we interpret as associated to outflowing gas. Conversely, no emission on the red side of \co32\ is detected at $v>600$ \kms.
The spatial resolution of our ALMA observation allows us to map the morphology of the outflow in extreme detail, as shown in Fig. \ref{fig:mappa-broad}a.   
Specifically,  the outflow in \pds\ shows several components: a bright inner component located at a radial distance $R\lesssim1.2$ kpc, plus an extended  component consisting of several clumps with different blueshifted bulk velocities located at radii $R\sim1.8-5$ kpc. 

\subsection{Extended outflow}

Fig. \ref{fig:mappa-broad}a shows the velocity integrated map of the \co32\ clumps in the $v\in[-1000,-250]$ \kms, obtained by integrating the emission detected at $>3\sigma$ in each 30 \kms\ spectral channel, for at least four contiguous channels (i.e. over a velocity range $\geq$ 120 \kms). The colour map shows that the outflowing gas is distributed in several clumps located at different distances from the QSO, up to $\sim5$ kpc, while the white contours refer to the quiescent molecular gas traced by the \co32\ core.
Two examples of the clumps spectra are given in Fig. \ref{fig:mappa-broad}b, showing that each clump emits over a typical range of $\sim200$ \kms. Specifically, the \co32\ spectrum at the position of clump A (located at $\sim0.6$ arcsec = 1.8 kpc from the nucleus) is characterised by an emission component centred at $v\sim0$, which is associated with the quiescent gas. It also shows an excess of blue-shifted emission at $v\sim-300$ and $v\sim-750$ \kms. Differently, the spectrum of clump B at a larger separation ($\sim1.6$ arcsec = 5 kpc) lacks systemic emission but shows a blue-shifted component centred at $v\sim-350$ \kms.
We model the spectrum of each extended clump with multiple Gaussian components (also shown in Fig. \ref{fig:mappa-broad}b).

\begin{figure}[htb]
    \centering
    \includegraphics[width=1\columnwidth]{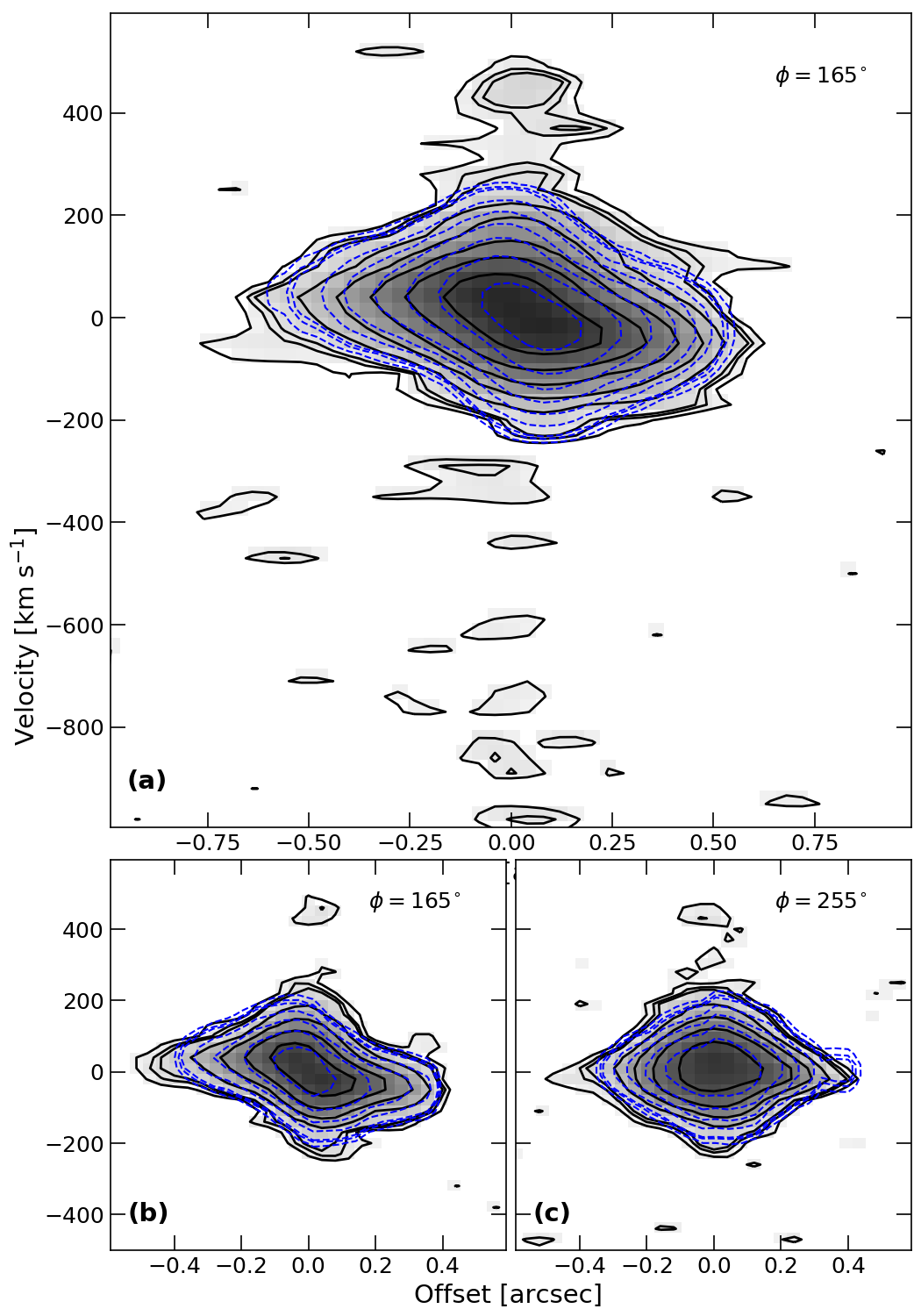}
    \caption{Panel \textit{\textbf{(a)}} shows the position-velocity diagram, extracted from a 0.3 arcsec slit centred on the QSO location along the major kinematic axis (see Fig. \ref{fig:core-moments}a), corresponding to a PA of 165 deg (measured anti-clockwise from north). Black contours refer to the [2, 3, 4,...2$^n$]$\sigma$ significance level of the \co32\ emission, with $\sigma = 0.083$ mJy beam$^{-1}$ and $n>2$. The contours associated with the best-fit $^{\rm 3D}$BAROLO model of the kinematics are also shown by the blue contours. Panels \textit{\textbf{(b)}} and \textit{\textbf{(c)}} are a zoom-in of the velocity range $v\in[-500,+600]$ \kms\ with increased angular resolution ($0.16\times0.19$ arcsec$^2$), extracted along and perpendicular to the major kinematic axis, respectively. Contours are as in top panel, with $\sigma = 0.16$ mJy beam$^{-1}$.}
    \label{fig:pv}
\end{figure}

\begin{figure}[htb]
    \centering
     \includegraphics[width = 1\columnwidth]{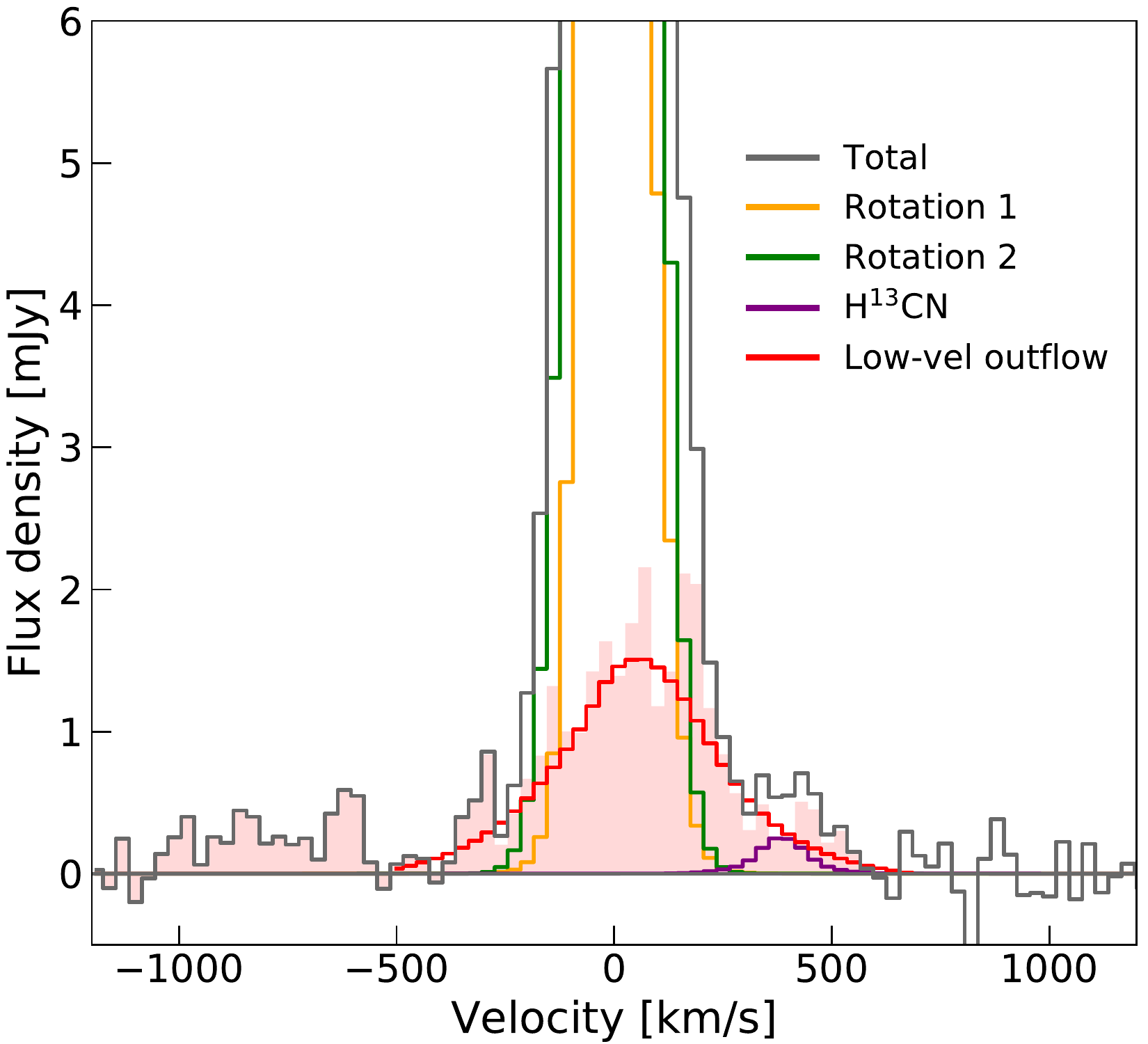}
    \caption{\co32\ spectrum of the central $1\times1$ arcsec$^2$ region centred on \pds. Data are shown in grey together with the total best-fit model resulting from the pixel-by-pixel decomposition in the $v\in[-500,+600$] \kms\ range, obtained by adding the best-fit models of each pixel. Yellow and green histograms indicate the two rotation components used to model the \co32\ core, while the emission from H$^{13}$CN at $v\sim+390$ \kms\ is shown in purple. The red histogram represents the best-fit of the low-velocity outflow component. The red shaded area represents the emission, remaining  after the subtraction of rotation and H$^{13}$CN emission, that we consider to be associated with the outflow, and indicates the presence of a blue-shifted, high-velocity ($v\sim-800$ \kms) outflow component.}
    \label{fig:decomposition}
\end{figure}

For a single high-velocity clump the mass outflow rate is computed as:
\begin{equation}\label{eq:mdot}
\dot{M}_{\rm mol} = \frac{M_{\rm mol}^{\rm out}\times v_{\rm 98}}{R}
\end{equation}

where $M_{\rm mol}^{\rm out}$ is the clump mass,  $v_{\rm 98}$ is the velocity enclosing 98\% of the cumulative velocity distribution of the outflowing gas, and $R$ is the projected distance of the clump from the QSO. 
$M_{\rm mol}^{\rm out}$ is derived from the \co32\ luminosity of the clumps detected at $>3\sigma$ in the velocity range $v\in[-1000,-250]$ \kms. We use the $L^\prime$CO$_{3-2}/$L$^\prime$CO$_{1-0}$ luminosity ratio measured for the systemic emission, and a luminosity-to-gas-mass conversion factor $\alpha_{\rm CO} = 0.8$ \msun\ (K \kms\ pc$^2$)$^{-1}$, typical of star-forming QSO hosts (see Sect. 4 for further details).
By adding together the contribution of all clumps, we estimate the molecular gas mass, the molecular mass outflow rate and momentum flux of the extended outflow detected in \pds, see Table \ref{tab:outflow_prop}.

\subsection{Central outflow}\label{sect:central}

Fig. \ref{fig:core-moments} shows the velocity and velocity dispersion map of \co32\ emission of the inner 1 arcsec$^2$ region.  The latter is resolved in about 12 independent beams which allow us to detect a projected velocity gradient in approximately the  north-west to south direction with a rather small total range ($v\in[-50,+40]$ \kms). Emission with a flat velocity distribution around $v=0$ \kms\ smears the velocity gradient in an arc-like region extending from the QSO position to $\sim[+0.3,-0.3]$ arcsec. The maximum of the velocity dispersion is observed in the central region ($\sigma_{\rm vel}\sim80$ \kms), where beam-smearing effects are more prominent \citep{Davies11,Tacconi13}. A more reliable estimate of $\sigma_{\rm vel}$ is provided by the average $\sigma_{\rm vel}\sim40-50$ \kms in an annulus with $0.2<R<0.4$ arcsec. 
The kinematics in the central region of \pds\ is more complex than that of a rotating disk, as further supported by the position-velocity diagram shown in Fig. \ref{fig:pv}a, extracted along the maximum velocity gradient direction from a 0.3 arcsec slit. A rotation pattern can be identified, with a velocity gradient $\Delta v_{\rm blue-red}\sim200$ \kms, which is modified by the presence of an excess of emission due to gas with velocity $v\in[-1000,+600]$ \kms\ roughly centred at the position of the QSO. This appears more evidently in Fig. \ref{fig:pv}b,c showing zoom-in position-velocity diagrams of the $v\in[-500,+600]$ \kms\ velocity range with an increased angular resolution of $0.16\times0.19$ arcsec$^2$ (see Sect. \ref{sect:datared}), extracted along and perpendicular to the major kinematic axis direction, respectively. 

We fit a 3D tilted-ring model with $^{\rm 3D}$BAROLO \citep{DiTeodoro15} to the data to provide a zero order description of the kinematics. We exclude from the fit the region with an angular separation $\lesssim0.15$ arcsec from the nucleus, where the high-velocity gas perturbs the kinematics. This results into an inclination $i=25\pm10$ deg, being consistent with the value of $\sim25$ deg derived from the projected axes ratio, and an intrinsic circular velocity $v_{\rm rot}=1.3\times\Delta v_{\rm blue-red}/(2$sin$i)\sim280$ \kms\ \citep[e.g.][]{Tacconi13}. The implied virial dynamical mass is $M_{\rm dyn}={Dv_{\rm rot}^2/2G}\sim1.0\times10^{10}$ \msun, where $D\sim1.3$ kpc is the source size estimated as 1.5$\times$ the deconvolved major axis of the \co32\ emission. A comparable value, i.e. $M_{\rm dyn}(i = 25\ \rm deg) \sim 1.2\times10^{10}$ \msun\ is derived by using the relation $M_{\rm dyn} = 1.16\times10^{5}\times(0.75\times {\rm FWHM}/{\rm sin}i)^2 \times D$ \citep{Wang13,Bischetti18}. Using the inferred dynamical mass we derive an escape velocity from the central 1.3 kpc of \pds\ of $\sim280$ \kms.
 
By subtracting the $^{\rm 3D}$BAROLO model to the ALMA cube we find that strong ($\sim8$ \% of the total \co32 flux) positive residuals are present in the velocity range $v\in[-500,+600]$ \kms. 
It is likely that these residuals are due to an inner emission component associated with the outflow described in Sect. 3.1.
Therefore, we perform an accurate modelling of the spectrum of the central region to better disentangle the contribution provided by the outflow from the total \co32\ emission. 
Specifically, we use a pixel-by-pixel spectral decomposition in the range $v\in[-500,+600]$ \kms\ with a combination of four Gaussian components to model:
(a) the disk rotation (two components\footnote{The normalisation of the first component is initially set to the peak of the emission in each pixel, while that of second component is  set to be 1/10 of the first one.}, needed to account for the partially resolved velocity distribution, i.e. nearby emitting regions with different rotation velocities within the ALMA beam);
(b) the outflow (one component with $\sigma>90$ \kms, i.e. the maximum value measured in the velocity dispersion map of the \co32\ emission); 
(c) the possible contamination by H$^{13}$CN(4$-$3)  emission (rest frequency $\nu_{\rm rest} = 345.34$ GHz) to the red wing of the \co32\ emission line \citep{Sakamoto10}. This component has a fixed velocity offset of $+390$ \kms, corresponding to the spectral separation between H$^{13}$CN and \co32, and line width equal or larger than that of the main rotation component.
Fig. \ref{fig:decomposition} shows the spectrum of  the $1\times1$ arcsec$^2$ central region with the different components  obtained by adding together the best-fit models obtained from the pixel-by-pixel fit. 

We then subtract from the total spectrum the components due to disk rotation and H$^{13}$CN. The residuals (red histogram in Fig. \ref{fig:decomposition}) may be associated with emission from outflowing gas. It is worth noting that a spectral decomposition without the outflow component (i.e. maximising the contribution from the emission due to rotation) is able to account for at most $\sim$50\% of these residuals.

\begin{figure*}[htb]
	\centering
	\includegraphics[width = 1\textwidth]{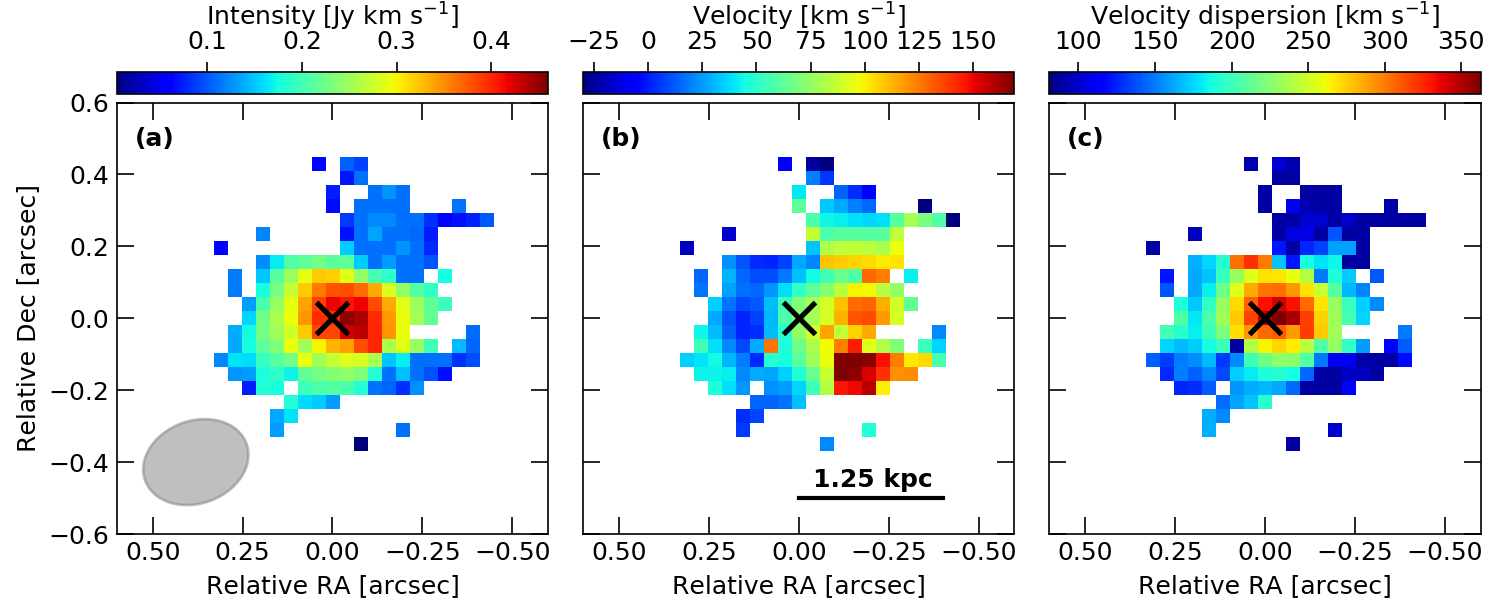}
	\caption{Intensity \textit{\textbf{(a)}}, velocity \textit{\textbf{(b)}} and velocity dispersion \textit{\textbf{(c)}} maps of the central, low-velocity outflow component resulting from the pixel-by-pixel decomposition of the \co32\ spectrum in the velocity range $v\in[-500,+600]$ \kms\ (red histogram in Fig. \ref{fig:decomposition}). The ALMA beam is shown by the grey ellipse, while the black cross indicates the QSO position.  }
	\label{fig:outflow-moments}
\end{figure*}

The bulk of this emission is due to a low velocity ($|v|\lesssim500$ \kms) component. Maps of the integrated flux density, velocity and velocity dispersion of this low-velocity emission component are shown in Fig. \ref{fig:outflow-moments}. This emission peaks at an offset of $\sim0.05$ arcsec ($\sim160$ pc) west from the QSO position (marked by a cross).
After deconvolving from the beam, the low-velocity outflow has a total projected physical scale of $\sim2.4$ kpc. A fraction of $\sim40$ \% of this emission is unresolved in the present ALMA observation.
A velocity gradient is detected along the east-west direction (see Fig. \ref{fig:outflow-moments}b), i.e. roughly perpendicular to the north-south gradient in the velocity map of the total \co32\ emission (see Fig. \ref{fig:core-moments}a). 
This molecular gas is also characterised by a high velocity dispersion (see Fig. \ref{fig:outflow-moments}c), with a peak value of $\sigma\sim360$ \kms\ and an average \citep{Davies11, Tacconi13} $\sigma\sim200$ \kms, suggesting highly turbulent gas close to the nucleus. 
All these pieces of evidence, in combination with the position-velocity diagram shown in Fig. \ref{fig:pv}, strongly suggest the presence of molecular gas whose kinematics is associated with outflowing gas. 

Beyond the velocity range $v\in[-500,+600]$ \kms\ covered by the spectral decomposition mentioned above, the \co32\ spectrum of the central $1\times1$ arcsec$^2$ region exhibits an excess of blue-shifted emission between $-500$ and $-1000$ \kms\ (see Fig. \ref{fig:decomposition}).
This high-velocity component can be modelled with a Gaussian line centred at $-800\pm80$ \kms, with flux density $0.25\pm0.08$ mJy and $\sigma=180\pm70$ \kms, and is visible in Fig. \ref{fig:mappa-broad}a at the position of the QSO. Based on its large velocity, this emission can be also associated to the molecular outflow in \pds. 

Accordingly, the red shaded area in Fig. \ref{fig:decomposition} represents the combination of the low- and high-velocity components for which we measure the outflow parameters (i.e. outflow mass, mass outflow rate and momentum flux) listed in Table \ref{tab:outflow_prop}. 
To avoid any possible contamination from H$^{13}$CN we exclude the spectral region in the range $v\in[310,560]$ \kms. 

As the central outflow is only marginally resolved by our ALMA observation, we infer its \mdotm\ by considering the simple scenario of a spherically/biconically symmetric, mass-conserving flow with constant velocity and uniform density up to $R\sim1.2$ kpc (Fig. \ref{fig:outflow-moments}), similarly to the geometry assumed for the molecular outflows detected in other luminous AGN, i.e. \citet{Vayner17,Feruglio17,Brusa18}. This corresponds to multiplying by a factor of three the \mdotm\ value inferred by Eq. \ref{eq:mdot}.  
Alternative outflow models considering a time-averaged thin shell geometry \citep[e.g.][]{Cicone15,Veilleux17} or a density profile scaling as $R^{-2}$ \citep{Rupke05} predict instead a mass outflow rate consistent with the value of \mdotm\ derived by Eq. \ref{eq:mdot}.

\begin{table*}[htb]
    \centering
    \setlength{\tabcolsep}{3pt}
    \caption{Main parameters of the molecular outflow detected in \pds. The outflowing gas mass, the mass outflow rate and momentum flux of the outflow (computed by using an $\alpha_{\rm CO}=0.8$ \msun(K \kms\ pc$^2$)$^{-1}$) are indicated for the different outflow components identified in the data. Brackets indicate the range of variation of each parameter considering 1$\sigma$ statistical errors and systematics associated with the spectral decomposition and uncertainty in the outflow physical size.
    }
    \begin{tabular}{lcccccccc}
    \toprule
      Outflow component & $R$ &  & \mout\  & \mdotm\ & \pdotm\  \\
      &  [kpc] & & [$10^8$ \msun] & [\msunyr] & [10$^{35}$ cm g s$^{-1}$] \\
    \midrule
    Extended & 1.8$-$5 & & $0.78[0.72-0.84]$ & $50[45-55]$ & $2.1[1.9-2.3]$ \\
    Central & $\lesssim1.2$ & $\left\{
    \begin{tabular}{l}
      $v\in[-500,+650]$ km s$^{-1}$\\
      $v<-500$ km s$^{-1}$
    \end{tabular}\right.\kern-\nulldelimiterspace$  & \begin{tabular}{c} $1.5[0.74^a-1.7]$ \\ $0.21[0.15^a-0.27]$ \end{tabular} & \begin{tabular}{c} $180[90-530^b]$ \\ $60[40-180^b]$ \end{tabular} & \begin{tabular}{c} $5.5[2.8-16^b]$ \\ $4.4[3.1-13^b]$ \end{tabular} \\
    
    \midrule
    Total & & & $2.5[1.6-2.8]$ & $290[180-760]$ & $12.0[7.8-32]$\\
    \bottomrule
    \end{tabular}
    \flushleft
    \small
    $^{a}$ The lower limit on \mout\ is computed by adding the residuals of the subtraction of the best fit pixel-by-pixel spectral decomposition with a model including only disk rotation and  H$^{13}$CN emission from the total CO spectrum.\\ 
    $^{b}$ The upper limits on \mdotm\ and \pdotm\ are derived assuming that the unresolved fraction ($\sim40$ \%) of the central outflow component has a minimum size of 160 pc. This value corresponds to $\sim1/4$ of the mean beam axis and to the spatial offset measured between the peaks of central outflow and total \co32\ emission (Sect. \ref{sect:central}).

    \label{tab:outflow_prop}
\end{table*}

\begin{figure*}
    \centering
    \includegraphics[width = 1\textwidth]{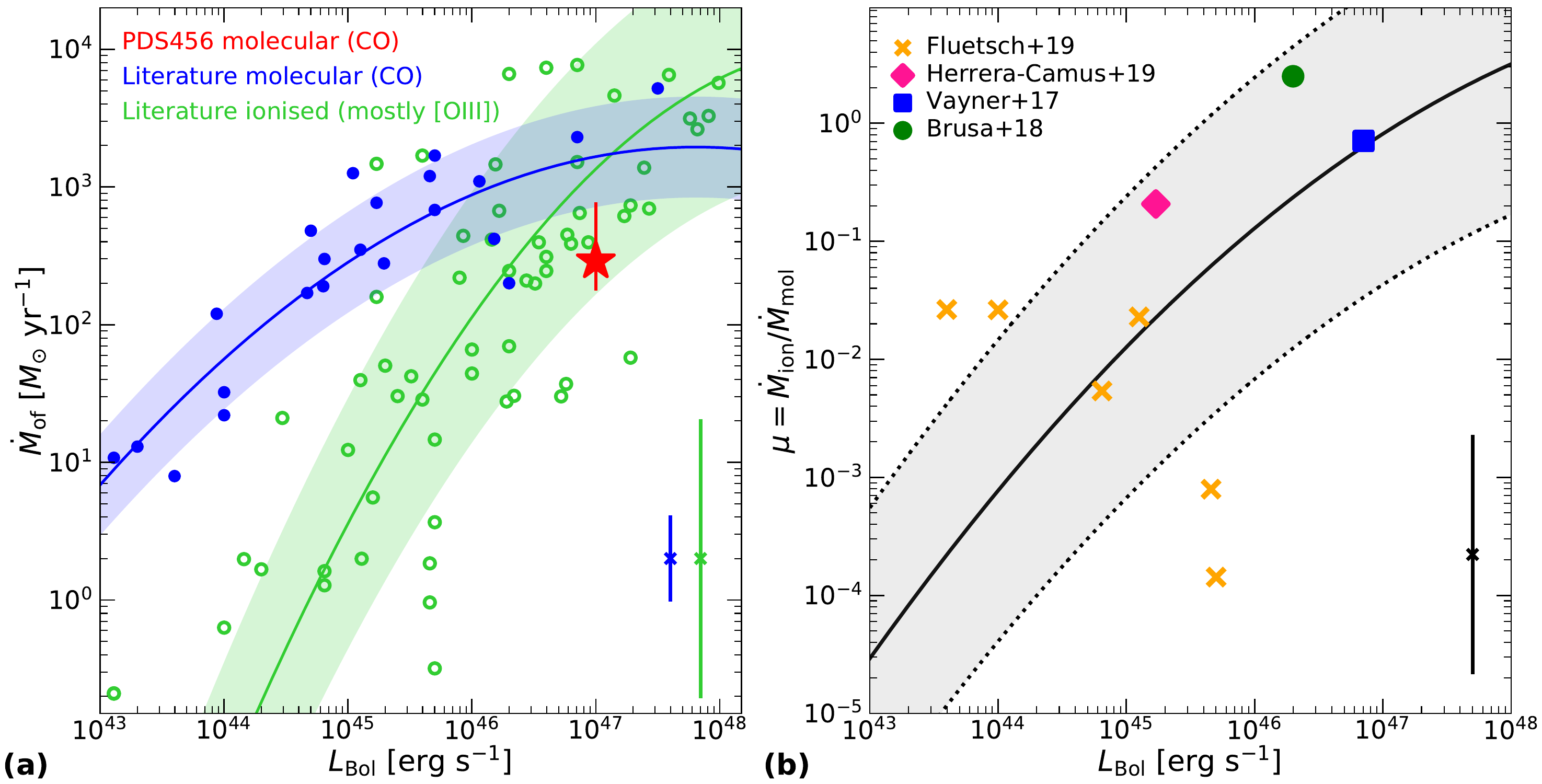}
    \caption{\textit{\textbf{(a):}} Mass outflow rate as a function of \lbol\ for \pds\ (red star) and a compilation of AGN with outflow detection from \citet{Fiore17} and \citet{Fluetsch19}, \citet{Zchaechner16, Feruglio17,Querejeta17,Vayner17,Veilleux17,Brusa18,Longinotti18,HerreraCamus19}. The blue(green) dashed line shows the best-fit parabolic function for the molecular(ionised) phase, while the shaded area indicates the rms scatter of the data from the relation. \textit{\textbf{(b):}} Ratio $\mu=\dot{M}_{\rm ion}/\dot{M}_{\rm mol}$ (black solid curve) inferred from the best-fit relations in panel (a),  as a function of \lbol. The shaded area represents the uncertainty on $\mu$, given the scatter of these relations.  Data points indicate the position in the $\mu-L_{\rm Bol}$ plane of the AGN with available molecular and ionised mass outflow rates \citep{Vayner17,Brusa18,HerreraCamus19}. We also include the sources with AGN contribution to \lbol\ $>10$ \% from \citet{Fluetsch19}. 
    }
    \label{fig:scalings}
\end{figure*}

\section{Discussion}

The ALMA observation of the \co32\ emission line in \pds\ reveals high-velocity molecular gas which traces a clumpy molecular outflow, extended out to $\sim$ 5 kpc from the nucleus in this hyper-luminous QSO. 
The molecular outflow discovered in \pds\ is the first reported for a radio quiet, non-lensed QSO in the poorly-explored brightest end ($L_{\rm Bol}\gtrsim10^{47}$ \ergs) of the AGN luminosity distribution.
The total mass of the outflowing molecular gas is \mout\ $\sim2.5\times 10^{8}$ \msun\ (for an $\alpha_{\rm CO}=0.8$ \msun(K \kms\ pc$^2$)$^{-1}$), of which $\sim$ 70 \% is located within the $\sim$ $2\times2$ kpc$^2$ inner region.
We stress that the high spatial resolution of our ALMA observation has been crucial to disentangle the outflow-related emission from the dominant emission of the quiescent gas.
The ratio between \mout\ and the total molecular gas mass for \pds\ is $\sim12$\%, which is comparable to ratios typically measured for other molecular outflows  \citep[$\sim10-20$\%, e.g.][]{Feruglio13a,Cicone14,Feruglio15,Brusa18}. We note that the estimate of the  molecular gas masses strongly depends on the assumption on $\alpha_{\rm CO}$. 
Given that (i) \pds\ exhibits a \lbol\ comparable to that of high-z QSOs with available CO measurements \citep{Carilli13} and (ii) the host-galaxy of \pds\ shows both a compact size and SFR comparable to local luminous infrared galaxies, we expect a $\alpha_{\rm CO}\sim0.8$ \msun (K \kms\ pc$^2$)$^{-1}$ in the central region, where the bulk of the outflowing gas lies \citep{Downes&Solomon93,Bolatto13}. Similarly to \citet{HerreraCamus19}, we adopt the same conversion factor also for the extended outflow. We note that \co32\ emission in the extended outflow clumps may turn optically thin because of the large velocity dispersion \citep{Bolatto13}. This would imply a lower  $\alpha_{\rm CO}\sim0.34$ \msun (K \kms\ pc$^2$)$^{-1}$ and, in turn, a smaller mass of the extended outflow by a factor of $\sim2.5$. On the other hand, assuming an $\alpha_{\rm CO}\sim2$ \msun (K \kms\ pc$^2$)$^{-1}$ as recently derived for the extended neutral outflow in NGC6240, i.e. a merging LIRG hosting two AGNs \citep{Cicone18}, would imply a larger total mass of the outflowing gas by a factor of $\sim2.5$.

By adding together the mass outflow rates of the inner and outer outflow components discovered by ALMA in \pds, we find a total \mdotm\ $\sim290$ \msunyr. This translates into a depletion timescale $\tau_{\rm dep} = M_{\rm mol}^{\rm out}/\dot{M}_{\rm mol}\sim8$ Myr for the molecular gas reservoir in \pds, suggesting a potential quenching of the star formation within a short time.
Such a $\tau_{\rm dep}$ is comparable to the Salpeter time for the mass growth rate of the SMBH in \pds\ and close to the typical QSO lifetime, indicating that this system will likely evolve into a passive galaxy with a dormant SMBH at its centre. 
Moreover, by including the measured rest-frame $\sim1$ mm continuum emission in a broad-band, UV-to-FIR fitting of the spectral energy distribution in \pds, we are able to measure a SFR$\sim30-80$ \msunyr\ in the QSO host galaxy (Vignali in prep.). In this process we avoid the contamination from the companions which account for the bulk of the FIR luminosity derived by previous, low resolution observations with only upper-limits in the FIR range above 100 micron \citep{Yun04}.

This implies that $\tau_{\rm dep}$ is a factor of $\sim4-10$ shorter than the time needed for the molecular gas to be converted into stars ($\tau_{\rm SF}$), indicating that the detected outflow is potentially able to affect the evolution of the host-galaxy in \pds.
A value of $\tau_{\rm dep}< \tau_{\rm SF}$  has been similarly observed for other molecular outflows observed in AGN  \citep[e.g.][]{Cicone14,Veilleux17,Brusa18,HerreraCamus19}. 
Given the large uncertainties both on $\tau_{\rm dep}$  and $\tau_{\rm SF}$, it is not possible to exclude a starburst contribution to the outflow acceleration unless $\tau_{\rm dep} << \tau_{\rm SF}$. However, for a given far-infrared luminosity, the estimate of the  SFR depends on the assumption  of the inital mass function (IMF) and star formation history. The SFR in the host-galaxy of \pds\ is estimated assuming a continuous star formation burst of $10-100$ Myr and a Salpeter IMF \citep{Kennicutt98}, in case of solar metallicity.  A different IMF (i.e. Larson or Chabrier) would translate into a smaller SFR by a factor of $\sim2$ \citep{Conroy09,Valiante14} and, hence, a larger $\tau_{\rm SF}$.


Fig. \ref{fig:scalings}a shows  the mass outflow rate \mdot\ as a function of \lbol\  for \pds\ and a compilation of molecular and ionised AGN-driven outflows from \citet{Fiore17}. We also include the molecular outflows recently revealed in CO emission by \citet{Zchaechner16, Feruglio17, Querejeta17, Veilleux17, Brusa18, Longinotti18} and those detected in both the molecular and ionised phase by \citet{Vayner17} and \citet{HerreraCamus19}, which have been identified to be AGN-driven. In addition to these outflows, we consider those reported by \citet{Fluetsch19} in systems where the AGN contributes to $\gtrsim20$ \% of \lbol, and that discovered by \citet{Pereira-Santaella18} in IRAS 14348$-$1447, for which the PA and the high ($\sim$ 800 \kms) velocity of the outflow suggest an AGN origin, for a total of 23(60) molecular(ionised) outflows. To minimise the systematic differences from sample to sample, all values have been recomputed from the tabulated data according to the same assumptions, following Eq. B.2 by \citet{Fiore17}. Nevertheless, some scatter between various samples may still be present because of the different assumptions in the literature on $\alpha_{\rm CO}$ and, therefore, on the outflow mass.

This updated compilation allows us to populate the luminosity range above $10^{46}$ \ergs, poorly sampled by both \citet{Fiore17} and \citet{Fluetsch19} samples.
The  relation for the molecular mass outflow rates as a function of \lbol\ by \cite{Fiore17} predicts values of \mdotm\ much larger than those measured for the sources with $L_{\rm Bol}>10^{46}$ \ergs. Accordingly, in order to model a likely flattening of the relation between the molecular mass outflow rate and \lbol\ in this high-luminosity range, we fit the molecular data with a parabolic function defined as:
\begin{equation}\label{eq:scaling}
    {\rm Log}\left(\frac{\dot{M}}{M_{\odot}}\right) = \alpha\times {\rm Log}^2\left(\frac{L_{\rm Bol}}{L_0}\right) + \beta\times {\rm Log}\left(\frac{L_{\rm Bol}}{L_0}\right) + \gamma 
\end{equation}

The best-fit relation is given by
$\alpha_{\rm mol} = -0.11$, $\beta_{\rm mol} = 0.80$, $\gamma_{\rm mol} = 1.78$ and $L_{\rm 0, mol}=10^{44.03}$ \ergs, with an associated scatter of $\sim0.37$ dex, computed as the rms of the molecular data points with respect to the relation. Our modelling suggests a molecular mass outflow rate $\dot{M}_{\rm mol}\sim1000$ \msunyr\ for  \lbol\ in the range $10^{46}-10^{48}$ \ergs. 
By fitting the ionised data with Eq. \ref{eq:scaling}, we find $\alpha_{\rm ion} = -0.21$, $\beta_{\rm ion} = 1.26$, $\gamma_{\rm ion} = 2.14$ and $L_{\rm 0, ion}=10^{46.07}$ \ergs, and a rms scatter of 0.91 dex. According to our best-fit relation, the ionised mass outflow rate $\dot{M}_{\rm ion}$ keeps increasing up to $L_{\rm Bol}\sim10^{48}$ \ergs.

\begin{figure*}[htb]
    \centering
    \includegraphics[width = 1\textwidth]{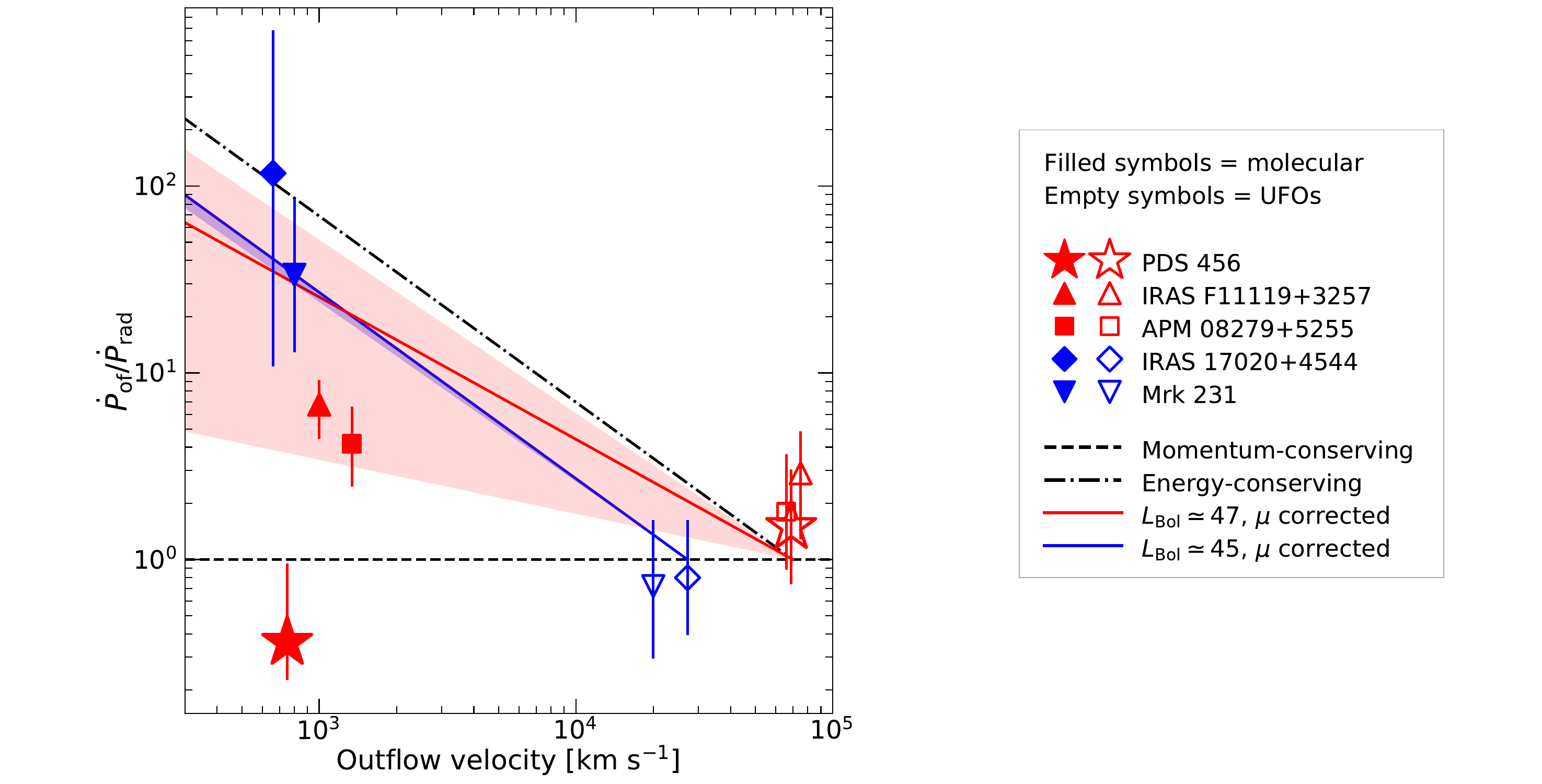}
    \caption{Ratio between the outflow (molecular or UFO) momentum flux and the radiative momentum flux as a function of the outflow velocity. Star = \pds; 
    Blue symbols = AGN with \lbol$<10^{46}$ \ergs\ \citep{Feruglio15, Longinotti18}; red symbols = AGN with \lbol$\sim10^{46}-10^{48}$ \citep[][and this work]{Tombesi15,Veilleux17,Feruglio17}.
    Filled symbols = molecular outflows; open symbols = UFOs. The dashed line is the expectation for a momentum-driven outflow. The dot-dashed line represents the prediction for an energy-driven outflow  with $\dot{P}_{\rm mol}/\dot{P}_{\rm rad} = v_{\rm UFO}/v$. The solid lines show the expected $\dot{P}_{\rm mol}/\dot{P}_{\rm rad}$ for a given luminosity and for different $\mu(L) = \dot{M}_{\rm ion}/\dot{M}_{\rm mol}$. The red(blue) shaded area shows the uncertainty on $\dot{P}_{\rm mol}/\dot{P}_{\rm rad}$ at $L_{\rm Bol}\sim10^{47}(10^{45})$ \ergs, given the rms scatter on $\dot{M}_{\rm ion}$ according to Eq. \ref{eq:scaling}.}
    \label{fig:energetics}
\end{figure*}

Fig. \ref{fig:scalings}b shows the ratio between the two parabolic functions described above which reproduce the ionised and molecular mass outflow rate trends with \lbol, i.e.  $\mu(L_{\rm Bol})=\dot{M}_{\rm ion}/\dot{M}_{\rm mol}$, in the luminosity range $L_{\rm Bol}\in[10^{43}-10^{48}]$ \ergs. Similarly to what previously noted by \citet{Fiore17}, we find that, although with a large scatter of about one order of magnitude, the ratio $\mu(L_{\rm Bol})=\dot{M}_{\rm ion}/\dot{M}_{\rm mol}$ increases with \lbol. The mean expected value varies between $\mu\sim10^{-3}$  at \lbol\ $\sim10^{44}$ \ergs\ and $\mu\sim1$ at \lbol\ $\sim10^{47}$ \ergs, suggesting a comparable contribution of the molecular and ionised gas phase to the outflow in \pds. 
In our analysis we do not take into account the contribution of the neutral atomic gas phase to the total mass outflow rate. However, for the few moderate-luminosity AGN with spatially-resolved measurements of the outflow in both the molecular and neutral gas phase, the latter seems to represent a fraction $\lesssim 30\%$ of the molecular one \citep[e.g.][]{Rupke&Veilleux13, Rupke17, Brusa18}.
In Fig. \ref{fig:scalings}b the sources with combined detection of outflow in the ionised and molecular gas phases are also shown. Differently from \citet{Fluetsch19}, who explored the luminosity range $L_{\rm Bol}\in[10^{44}-10^{46}]$ \ergs\ finding an anti-correlation between $\mu$ and \lbol, we observe that a positive trend is likely present  over a wider luminosity range, although the limited statistics \citep[three more objects with respect to][]{Fluetsch19} does not allow to put any firm conclusion on this.
However, we note that values of $\mu$ close or even larger than unity have already been reported for moderately-luminous QSOs \citep[e.g.][]{Vayner17,Brusa18}, suggesting that sources with comparable molecular and ionised mass outflow rates may span a wide range of \lbol.

In Fig. \ref{fig:energetics} we plot the outflow momentum boost \pdot/$\dot{P}_{\rm rad}$, where $\dot{P}_{\rm rad}=L_{\rm Bol}/c$, as a function of the outflow velocity\footnote{Similarly to what has been done for the molecular outflows, all $\dot{P}_{\rm UFO}$ have been homogeneously recomputed according to Eq. 2 in \citet{Nardini18}.}. 
This plot has been often used to compare different models of energy transfer between UFOs and galaxy-scale outflows, assuming that most of the outflow mass is carried by the molecular phase, i.e. \pdot\ $\sim$ \pdotm\ \citep{Tombesi15,Feruglio15}. This may not be true, especially in the high \lbol\ regime, as in the case of  \pds\ (Fig. \ref{fig:energetics}a). 
The ratio \pdotm/$\dot{P}_{\rm rad}$ $\sim0.36$ estimated using CO for the galactic scale outflow in \pds\ is significantly smaller than those measured in other AGN, typically showing  \pdotm/$\dot{P}_{\rm rad}$ $\sim$ $5-50$. Interestingly, it is of the order of $\dot{P}_{\rm UFO}$/$\dot{P}_{\rm rad}$, found by \citet{Nardini15,Luminari18}.


The dot-dashed line indicates the expected \pdot/$P_{\rm rad}$ for an energy conserving expansion assuming that most of the outflow mass is traced by the molecular phase. As suggested by Fig. \ref{fig:scalings}, this is likely not the case in the high \lbol\ regime, where the ionised outflow can be as massive as the molecular one. 
We thus probably detect in the molecular phase only a fraction of the total outflowing mass in \pds. Therefore, when comparing the expectation for the energy-conserving scenario with the results of ALMA observations we need to take into account that using the molecular phase alone to estimate the outflow mass may lead to underestimate the total mass of the outflow (i.e. the y-position of the red star marking \pds\ in Fig. \ref{fig:energetics} should be considered as a lower limit). We thus use an empirical relation to estimate the molecular momentum flux $\dot{P}_{\rm mol}$ using the scaling relations given by Eq. \ref{eq:scaling}. 
Specifically, the ratio between the total momentum flux of the large-scale outflow and that of the UFO for an energy-conserving expansion is related to the UFO and outflow velocities ($v_{\rm UFO}$ and $v_{\rm of}$ ) by the following relation:
\begin{equation}
\frac{\dot{P_{\rm of}}}{\dot{P_{\rm UFO}}} = \frac{v_{\rm UFO}}{v_{\rm of}}
\end{equation}
that, by assuming $\dot{P}_{\rm of}\sim\dot{P}_{\rm mol}+\dot{P}_{\rm ion}$, translates into a ratio $\dot{P}_{\rm mol}/\dot{P}_{\rm rad}$ given by:
\begin{equation}\label{eq:pdot}
\frac{\dot{P}_{\rm mol}}{\dot{P}_{\rm rad}} = \frac{v_{\rm UFO}}{v_{\rm mol}}\times\frac{1}{1+k\times\mu(L_{\rm Bol})}
\end{equation}
where $v_{\rm mol}$ is the velocity of the molecular and outflow and $k = v_{\rm ion}/v_{\rm mol}$ is the ratio between the velocity of the ionised outflow and $v_{\rm mol}$. For our calculations, we assume $v_{\rm mol}\sim1000$ \kms\ and $k\sim2$ \citep{Fiore17}.

Solid lines plotted in Fig. \ref{fig:energetics} represent the relations inferred from Eq. \ref{eq:pdot} for a luminosity of $L_{\rm Bol}\sim10^{45}$ and $\sim10^{47}$ \ergs, respectively. 
We note that for AGN at relatively low luminosity (such as Mrk 231 and IRAS 17020$+$4544) the relation has a similar slope to the classic energy-conserving model, for which $\mu(L_{\rm Bol})<<1$, because the bulk of the outflowing mass is due to molecular gas. Conversely, for hyper-luminous AGN, the empirical relation for $\dot{P}_{\rm mol}/\dot{P}_{\rm rad}$ is less steep, as expected when $\mu(L_{\rm Bol})$ increases. This effect reduces the discrepancy between the observed  $\dot{P}_{\rm mol}/\dot{P}_{\rm rad}$ and the expectation for a "luminosity-corrected" energy-conserving scenario. 

So far, there is no available observation of the outflow in the ionised gas phase for these hyper-luminous sources. However, it is interesting to note that a massive ionised outflow characterised by a $\dot{M}_{\rm ion}\gtrsim10^3$ \msunyr, as inferred from Eq. \ref{eq:scaling} at such high luminosities, would be required to fit the measured $\dot{P}_{\rm mol}/\dot{P}_{\rm rad}$ in IRAS F11119$+$3257 and APM 08279$+$5255 (red shaded area in Fig. \ref{fig:energetics}). Remarkably, in the case of \pds, even a $\dot{M}_{\rm ion}$ as large as $10^4$ \msunyr\ (i.e. the maximum  value allowed for an ionised outflow by Eq. \ref{eq:scaling} and its associated scatter) would likely be still insufficient to match the expectation for an energy-conserving outflow.
On the other hand, the small value of the momentum boost derived for \pds\ may be an indication that the shocked gas by the UFO  preferentially expands along a direction not aligned with the disc plane and is not able to sweep up large amount of ISM (Menci et al. 2019, submitted).


Alternatively, the results of our analysis can be interpreted as an indication of forms of outflow driving mechanisms in high-luminosity AGN different
from the UFO-related energy-driving.
Models based on a mechanism for driving galaxy-scale outflows via radiation-pressure on dust indeed predict
$\dot{P}_{\rm mol}/\dot{P}_{\rm rad}$ values around unity, and may offer a viable explanation for the observed energetics of the outflow in \pds\ \citep{Ishibashi&Fabian14,Thompson15,Costa18a, Ishibashi18}.
On the other hand, large-scale ($\gtrsim$ a few hundreds of pc) outflows cannot be explained by a momentum-conserving expansion which predicts  a rapid cooling of the shocked wind \citep[e.g.][]{King&Pounds15}.

\begin{figure}
    \centering
    \includegraphics[width=1\columnwidth]{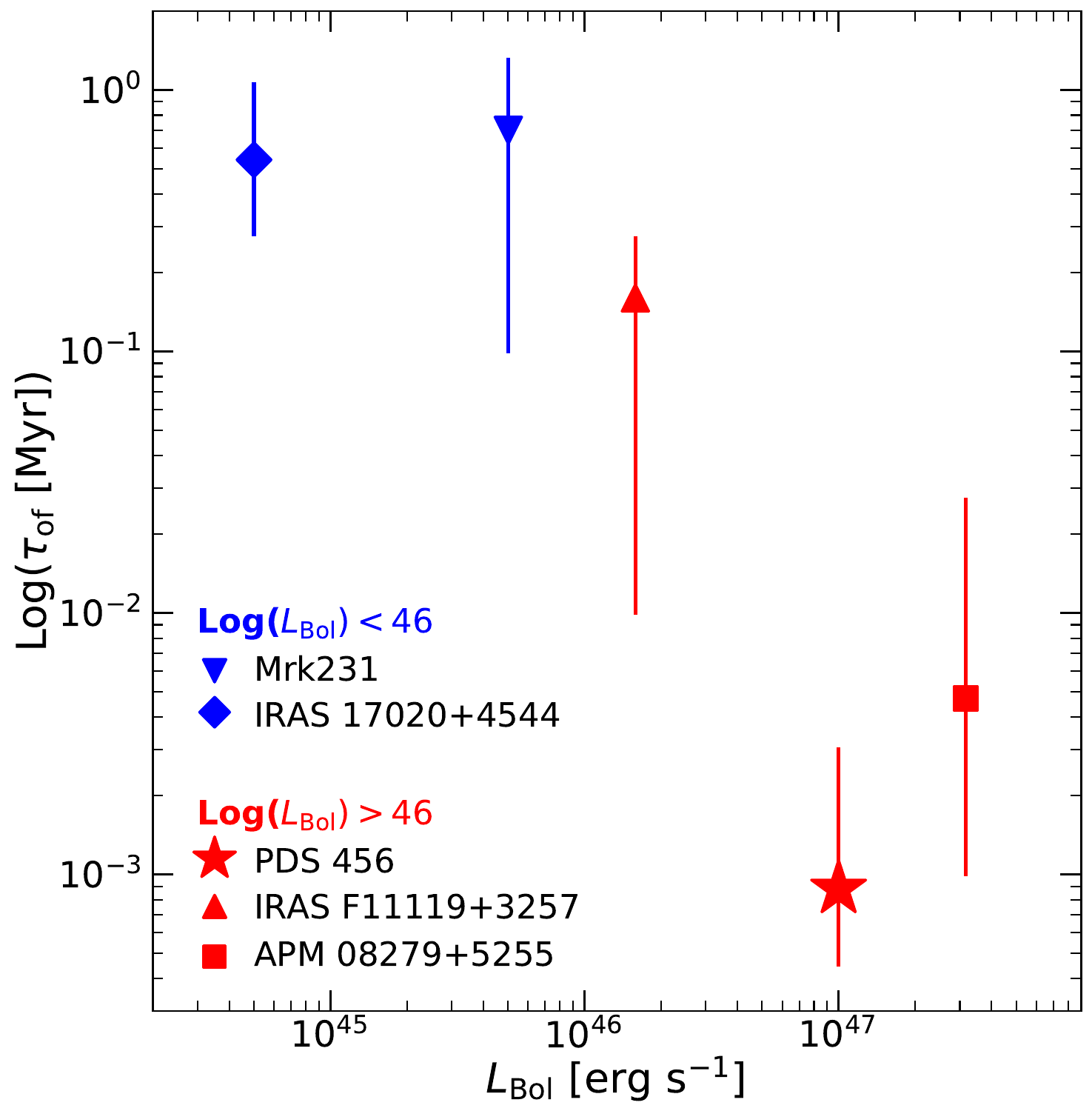}
    \caption{$\tau_{\rm of}=E_{\rm of}/\dot{E}_{\rm UFO}$ as a function of \lbol\ for \pds\ and a compilation of AGN as in Fig \ref{fig:energetics}. $E_{\rm of}$ is computed under the assumption that the bulk of the outflowing gas mass is in the molecular phase. Blue(red) symbols correspond to AGN with $L_{\rm Bol}<10^{46}$ \ergs\ ($L_{\rm Bol}>10^{46}$ \ergs).}
    \label{fig:tau}
\end{figure}

Fig. \ref{fig:tau} shows $\tau_{\rm of} = E_{\rm of}/\dot{E}_{\rm UFO}$, which represents the time needed for the relativistic wind to provide the mechanical energy of the galaxy-scale outflow integrated over its flow time, i.e. $E_{\rm of}=0.5\times M_{\rm of}v_{\rm of}^2$, as a function of \lbol.
Being a function of $E_{\rm of}$, $\tau_{\rm of}$ allows to constrain the UFO efficiency in producing the observed kpc-scale outflow without any assumption on its morphology and size \citep{Nardini18}. 
For AGN with $L_{\rm Bol}<10^{46}$ \ergs\ $\tau_{\rm of}\sim10^5-10^6$ yrs while it drops to $\sim10^3$ yrs in hyper-luminous QSOs such as \pds, suggesting a high efficiency of the UFO launched in these sources.
We note that $E_{\rm of}$ should in principle be derived by including all gas phases at a given radius \citep{Nardini18}, while in Fig. \ref{fig:tau} we only consider the molecular gas. In fact, complementary information on both the molecular and ionised phases as traced by e.g. CO and [O III] is typically unavailable.
 Therefore, the observed trend of a decreasing $\tau_{\rm of}$ with increasing \lbol\ may be a further indication of a smaller molecular gas contribution to the total energy carried by the outflow in the high luminosity regime. However, since $\tau_{\rm of}$ is the ratio between a time-averaged quantity ($E_{\rm of}$) and an instantaneous quantity ($\dot{E}_{\rm UFO}$), a small value may be also explained in terms of an "outburst" phase of the UFO in the two sources with $L_{\rm Bol}\sim10^{47}$ \ergs\ considered here (i.e., PDS 456 and APM 08279$+$5255).
Alternatively, a small coupling of the UFO with the host-galaxy ISM can be invoked to account for the short $\tau_{\rm of}$ observed in these QSOs in a scenario where the kpc-scale outflow is undergoing an energy conserving expansion. 


\section{Summary and Conclusions}
In this work, we report on the ALMA observation of the 1~mm continuum and \co32\ line emission in \pds\ ($z_{\rm CO}=0.185$). These data enable us to probe with unprecedented spatial resolution ($\sim700$ pc) the ISM in the host-galaxy of a hyper-luminous ($L_{\rm Bol}\sim10^{47}$ \ergs) QSO. We provide the first detection of a molecular outflow in a radio-quiet, non-lensed source at the brightest end of the AGN luminosity function. Our observation highlights the importance of the combined high spatial resolution and high sensitivity provided by ALMA in revealing broad wings much weaker than the core of the CO emission line, and disentangling the relative contribution of outflowing and quiescent molecular gas to the emission from the innermost regions around QSOs. Our main findings can be summarised as follows:
\begin{itemize}
    \item We detect at $\sim350\sigma$ significance the \co32\ emission from the host-galaxy of \pds, finding that the bulk of the molecular gas reservoir is located in a rotating disk with compact size ($\sim1.3$ kpc) seen under a small inclination ($i\sim25$ deg), with an intrinsic circular velocity of $\sim280$ \kms. We measure a molecular gas of $M_{\rm mol}\sim2.5\times10^9$ \msun\ and a dynamical mass of $M_{\rm dyn}\sim1\times10^{10}$ \msun.
    \item The \co32\ emission line profile shows a blue-shifted tail (whose flux density is about $1/60$ of the line peak), extending to $v\sim-1000$ \kms, and a red-shifted wing at $v\lesssim600$ \kms, associated with molecular outflowing gas. The outflow is characterised by a complex morphology, as several clumps with blue-shifted velocity are detected over a wide region out to $\sim5$ kpc from the nucleus, in addition to a bright compact, outflow component with velocity $v\in[-1000,500]$ \kms\ located within $\sim1.2$ kpc.
    \item By adding together all outflow components, we measure a total mass  \mout\ $\sim2.5\times10^8$ \msun\ and a mass outflow rate \mdotm\ $\sim290$ \msunyr. This is a remarkably weak outflow for such a hyper-luminous QSO hosting one of the fastest and most energetic UFO ever detected.
    Nevertheless, the measured \mdotm\ implies a depletion timescale $\tau_{\rm dep}\sim8$ Myr for the molecular gas in \pds, being a factor of 4 $-$ 10 shorter than the gas depletion time based on the SFR.
    This suggests a possible quenching of the star-formation activity in the host galaxy within a short time.
    \item The momentum boost of the molecular outflow, with respect to the AGN radiative momentum output is  \pdotm/$\dot{P}_{\rm rad}\sim0.36$, which represents the smallest value reported so far for sources exhibiting both UFO and molecular
    outflow. This result improves our understanding  of the \pdot/$\dot{P}_{\rm rad}$ versus \lbol\ relation and   indicates
    that the  relation between UFO and galaxy-scale molecular outflow
is very complex and may significantly differ from the typical expectations  of  models of  energy-conserving expansion 
\citep[e.g.][]{Faucher-Giguere12,Zubovas&King12}, i.e.  \pdot/$\dot{P}_{\rm rad}$ $\gg$ 1.

    \item We calculate updated scaling relations between the mass outflow rate and \lbol\ for both  the molecular and ionised gas phase. Thanks to our detection of the molecular outflow in \pds, combined with other recent results, we can extend the modelling of the  \mdotm\ vs \lbol\ relation by one order of magnitude in luminosity.
    Our best-fit relations indicate that the molecular mass outflow rate flattens at \lbol\ $>10^{46}$ \ergs, while the ionised one keeps increasing up to \lbol\ $\sim10^{48}$ \ergs. Although with a large scatter, the two gas phases appear comparable  at \lbol\ $\sim10^{47}$ \ergs, suggesting that in luminous QSOs the ionised gas phase cannot be neglected  to properly evaluate the impact of AGN-driven feedback.
    Planned high-resolution \textit{VLT}-MUSE observations will offer us an excellent opportunity to shed light on this by probing the ionised gas phase in \pds\ with unprecedented detail. 
    \item We derive an empirical relation to compute the luminosity-corrected \pdotm\ in case of an energy-conserving scenario, as a function of \lbol. Accordingly, we  predict smaller  $\dot{P}_{\rm mol}$/$\dot{P}_{\rm rad}$ in luminous QSOs compared to the "classic" energy-conserving scenario. However, in case of \pds, the smallest $\dot{P}_{\rm mol}$ predicted by our analysis (corresponding to a $\dot{M}_{\rm ion}\sim10^{4}$ \msunyr) still falls short on matching the expectations for an efficient energy-conserving expansion, unless the shocked gas by the UFO leaks out  along a direction that intercepts a small fraction of the molecular disc.
    Remarkably, the small momentum boost measured for the molecular outflow in \pds\ lends support to  a driving mechanism alternative to or concurrent with energy-driving, i.e.  the 
    AGN radiation pressure on dust, predicting momentum ratios close to unity. 
    
    \item The time necessary for the UFO to supply the energy measured for the molecular outflow in \pds, i.e. $\tau_{\rm of}\sim10^{-3}$ Myr, is about two orders of magnitude shorter than those derived for AGN at lower \lbol. Such a small value of $\tau_{\rm of}$ may suggest that the molecular phase is not representative of the total outflow energy in hyper-luminous sources, or that the UFO  in \pds\ is caught in an "outburst" phase. Alternatively, it may be an indication of AGN radiative feedback at work in luminous QSOs. All these hypotheses suggest a very complex interplay between nuclear activity and its surroundings,
    with important  implications for evaluating and simulating the impact and role of AGN-driven in the evolution of massive galaxies.  
    
\end{itemize}
\begin{acknowledgements}
This paper makes use of the ALMA data from project ADS/JAO.ALMA\#2016.1.01156.S. ALMA is a partnership of ESO (representing its member states), NSF (USA) and NINS (Japan), together with NRC (Canada), MOST and ASIAA (Taiwan), and KASI (Republic of Korea), in cooperation with the Republic of Chile. The Joint ALMA Observatory is operated by ESO, AUI/NRAO and NAOJ. We thank R. Valiante, C. Ramos-Almeida and N. Menci for helpful discussions, and E. Di Teodoro and M. Talia for his assistance in the usage of the $^{\rm 3D}$BAROLO model. 
M. Bischetti, E. Piconcelli, A. Bongiorno,  L. Zappacosta and M. Brusa acknowledge financial support from  ASI and INAF under the contract 2017-14-H.0 ASI-INAF. C. Feruglio, E. Piconcelli and F. Fiore acknowledge financial support from INAF under the contract PRIN INAF 2016 FORECAST. R. Maiolino acknowledges ERC Advanced Grant 695671 "QUENCH" and support by the Science and Technology Facilities Council (STFC). C. Cicone and E. Nardini acknowledge funding from the European Union's Horizon 2020 research and innovation program under the Marie Sklodowska-Curie grant agreement No 664931.
\end{acknowledgements}

\bibpunct{(}{)}{;}{a}{}{,} 
\bibliographystyle{aa} 
\bibliography{biblio} 

\begin{thebibliography}{73}
\expandafter\ifx\csname natexlab\endcsname\relax\def\natexlab#1{#1}\fi

\bibitem[{{Aalto} {et~al.}(2015){Aalto}, {Garcia-Burillo}, {Muller}, {Winters},
  {Gonzalez-Alfonso}, {van der Werf}, {Henkel}, {Costagliola}, \&
  {Neri}}]{Aalto15}
{Aalto}, S., {Garcia-Burillo}, S., {Muller}, S., {et~al.} 2015, \aap, 574, A85

\bibitem[{{Alatalo}(2015)}]{Alatalo15}
{Alatalo}, K. 2015, \apjl, 801, L17

\bibitem[{{Alatalo} {et~al.}(2015){Alatalo}, {Lacy}, {Lanz}, {Bitsakis},
  {Appleton}, {Nyland}, {Cales}, {Chang}, {Davis}, \& {de Zeeuw}}]{Alatalo15a}
{Alatalo}, K., {Lacy}, M., {Lanz}, L., {et~al.} 2015, \apj, 798, 31

\bibitem[{{Barcos-Mu{\~n}oz} {et~al.}(2018){Barcos-Mu{\~n}oz}, {Aalto},
  {Thompson}, {Sakamoto}, {Mart{\'\i}n}, {Leroy}, {Privon}, {Evans}, \&
  {Kepley}}]{BarcosMunoz18}
{Barcos-Mu{\~n}oz}, L., {Aalto}, S., {Thompson}, T.~A., {et~al.} 2018, \apj,
  853, L28

\bibitem[{{Bischetti} {et~al.}(2018){Bischetti}, {Piconcelli}, {Feruglio},
  {Duras}, {Bongiorno}, {Carniani}, {Marconi}, {Pappalardo}, {Schneider},
  {Travascio}, {Valiante}, {Vietri}, {Zappacosta}, \& {Fiore}}]{Bischetti18}
{Bischetti}, M., {Piconcelli}, E., {Feruglio}, C., {et~al.} 2018, \aap, 617,
  A82

\bibitem[{{Bischetti} {et~al.}(2017){Bischetti}, {Piconcelli}, {Vietri},
  {Bongiorno}, {Fiore}, {Sani}, {Marconi}, {Duras}, {Zappacosta}, {Brusa},
  {Comastri}, {Cresci}, {Feruglio}, {Giallongo}, {La Franca}, {Mainieri},
  {Mannucci}, {Martocchia}, {Ricci}, {Schneider}, {Testa}, \&
  {Vignali}}]{Bischetti17}
{Bischetti}, M., {Piconcelli}, E., {Vietri}, G., {et~al.} 2017, \aap, 598, A122

\bibitem[{{Bolatto} {et~al.}(2013){Bolatto}, {Wolfire}, \& {Leroy}}]{Bolatto13}
{Bolatto}, A.~D., {Wolfire}, M., \& {Leroy}, A.~K. 2013, \araa, 51, 207

\bibitem[{{Brusa} {et~al.}(2018){Brusa}, {Cresci}, {Daddi}, {Paladino},
  {Perna}, {Bongiorno}, {Lusso}, {Sargent}, {Casasola}, {Feruglio},
  {Fraternali}, {Georgiev}, {Mainieri}, {Carniani}, {Comastri}, {Duras},
  {Fiore}, {Mannucci}, {Marconi}, {Piconcelli}, {Zamorani}, {Gilli}, {La
  Franca}, {Lanzuisi}, {Lutz}, {Santini}, {Scoville}, {Vignali}, {Vito},
  {Rabien}, {Busoni}, \& {Bonaglia}}]{Brusa18}
{Brusa}, M., {Cresci}, G., {Daddi}, E., {et~al.} 2018, \aap, 612, A29

\bibitem[{{Bryant} \& {Scoville}(1999)}]{Bryant&Scoville99}
{Bryant}, P.~M. \& {Scoville}, N.~Z. 1999, \aj, 117, 2632

\bibitem[{{Carilli} \& {Walter}(2013)}]{Carilli13}
{Carilli}, C.~L. \& {Walter}, F. 2013, \araa, 51, 105

\bibitem[{{Choi} {et~al.}(2018){Choi}, {Somerville}, {Ostriker}, {Naab}, \&
  {Hirschmann}}]{Choi18}
{Choi}, E., {Somerville}, R.~S., {Ostriker}, J.~P., {Naab}, T., \&
  {Hirschmann}, M. 2018, \apj, 866, 91

\bibitem[{{Cicone} {et~al.}(2015){Cicone}, {Maiolino}, {Gallerani}, {Neri},
  {Ferrara}, {Sturm}, {Fiore}, {Piconcelli}, \& {Feruglio}}]{Cicone15}
{Cicone}, C., {Maiolino}, R., {Gallerani}, S., {et~al.} 2015, \aap, 574, A14

\bibitem[{{Cicone} {et~al.}(2014){Cicone}, {Maiolino}, {Sturm},
  {Graci{\'a}-Carpio}, {Feruglio}, {Neri}, {Aalto}, {Davies}, {Fiore},
  {Fischer}, {Garc{\'{\i}}a-Burillo}, {Gonz{\'a}lez-Alfonso},
  {Hailey-Dunsheath}, {Piconcelli}, \& {Veilleux}}]{Cicone14}
{Cicone}, C., {Maiolino}, R., {Sturm}, E., {et~al.} 2014, \aap, 562, A21

\bibitem[{{Cicone} {et~al.}(2018){Cicone}, {Severgnini}, {Papadopoulos},
  {Maiolino}, {Feruglio}, {Treister}, {Privon}, {Zhang}, {Della Ceca}, {Fiore},
  {Schawinski}, \& {Wagg}}]{Cicone18}
{Cicone}, C., {Severgnini}, P., {Papadopoulos}, P.~P., {et~al.} 2018, \apj,
  863, 143

\bibitem[{{Conroy} {et~al.}(2009){Conroy}, {Gunn}, \& {White}}]{Conroy09}
{Conroy}, C., {Gunn}, J.~E., \& {White}, M. 2009, \apj, 699, 486

\bibitem[{{Costa} {et~al.}(2018){Costa}, {Rosdahl}, {Sijacki}, \&
  {Haehnelt}}]{Costa18a}
{Costa}, T., {Rosdahl}, J., {Sijacki}, D., \& {Haehnelt}, M.~G. 2018, \mnras,
  473, 4197

\bibitem[{{Costa} {et~al.}(2014){Costa}, {Sijacki}, \& {Haehnelt}}]{Costa14}
{Costa}, T., {Sijacki}, D., \& {Haehnelt}, M.~G. 2014, \mnras, 444, 2355

\bibitem[{{Croton} {et~al.}(2006){Croton}, {Springel}, {White}, {De Lucia},
  {Frenk}, {Gao}, {Jenkins}, {Kauffmann}, {Navarro}, \& {Yoshida}}]{Croton06}
{Croton}, D.~J., {Springel}, V., {White}, S.~D.~M., {et~al.} 2006, \mnras, 365,
  11

\bibitem[{{Davies} {et~al.}(2011){Davies}, {F{\"o}rster Schreiber}, {Cresci},
  {Genzel}, {Bouch{\'e}}, {Burkert}, {Buschkamp}, {Genel}, {Hicks}, {Kurk},
  {Lutz}, {Newman}, {Shapiro}, {Sternberg}, {Tacconi}, \& {Wuyts}}]{Davies11}
{Davies}, R., {F{\"o}rster Schreiber}, N.~M., {Cresci}, G., {et~al.} 2011,
  \apj, 741, 69

\bibitem[{{Di Teodoro} \& {Fraternali}(2015)}]{DiTeodoro15}
{Di Teodoro}, E.~M. \& {Fraternali}, F. 2015, \mnras, 451, 3021

\bibitem[{{Downes} {et~al.}(1993){Downes}, {Solomon}, \&
  {Radford}}]{Downes&Solomon93}
{Downes}, D., {Solomon}, P.~M., \& {Radford}, S.~J.~E. 1993, \apjl, 414, L13

\bibitem[{{Evans} {et~al.}(2006){Evans}, {Solomon}, {Tacconi}, {Vavilkin}, \&
  {Downes}}]{Evans06}
{Evans}, A.~S., {Solomon}, P.~M., {Tacconi}, L.~J., {Vavilkin}, T., \&
  {Downes}, D. 2006, \aj, 132, 2398

\bibitem[{{Faucher-Gigu{\`e}re} \& {Quataert}(2012)}]{Faucher-Giguere12}
{Faucher-Gigu{\`e}re}, C.-A. \& {Quataert}, E. 2012, \mnras, 425, 605

\bibitem[{{Feruglio} {et~al.}(2017){Feruglio}, {Ferrara}, {Bischetti},
  {Downes}, {Neri}, {Ceccarelli}, {Cicone}, {Fiore}, {Gallerani}, {Maiolino},
  {Menci}, {Piconcelli}, {Vietri}, {Vignali}, \& {Zappacosta}}]{Feruglio17}
{Feruglio}, C., {Ferrara}, A., {Bischetti}, M., {et~al.} 2017, \aap, 608, A30

\bibitem[{{Feruglio} {et~al.}(2015){Feruglio}, {Fiore}, {Carniani},
  {Piconcelli}, {Zappacosta}, {Bongiorno}, {Cicone}, {Maiolino}, {Marconi},
  {Menci}, {Puccetti}, \& {Veilleux}}]{Feruglio15}
{Feruglio}, C., {Fiore}, F., {Carniani}, S., {et~al.} 2015, \aap, 583, A99

\bibitem[{{Feruglio} {et~al.}(2013{\natexlab{a}}){Feruglio}, {Fiore},
  {Maiolino}, {Piconcelli}, {Aussel}, {Elbaz}, {Le Floc'h}, {Sturm}, {Davies},
  \& {Cicone}}]{Feruglio13a}
{Feruglio}, C., {Fiore}, F., {Maiolino}, R., {et~al.} 2013{\natexlab{a}}, \aap,
  549, A51

\bibitem[{{Feruglio} {et~al.}(2013{\natexlab{b}}){Feruglio}, {Fiore},
  {Piconcelli}, {Cicone}, {Maiolino}, {Davies}, \& {Sturm}}]{Feruglio13b}
{Feruglio}, C., {Fiore}, F., {Piconcelli}, E., {et~al.} 2013{\natexlab{b}},
  \aap, 558, A87

\bibitem[{{Feruglio} {et~al.}(2010){Feruglio}, {Maiolino}, {Piconcelli},
  {Menci}, {Aussel}, {Lamastra}, \& {Fiore}}]{Feruglio10}
{Feruglio}, C., {Maiolino}, R., {Piconcelli}, E., {et~al.} 2010, \aap, 518,
  L155

\bibitem[{{Fiore} {et~al.}(2017){Fiore}, {Feruglio}, {Shankar}, {Bischetti},
  {Bongiorno}, {Brusa}, {Carniani}, {Cicone}, {Duras}, {Lamastra}, {Mainieri},
  {Marconi}, {Menci}, {Maiolino}, {Piconcelli}, {Vietri}, \&
  {Zappacosta}}]{Fiore17}
{Fiore}, F., {Feruglio}, C., {Shankar}, F., {et~al.} 2017, \aap, 601, A143

\bibitem[{{Fluetsch} {et~al.}(2019){Fluetsch}, {Maiolino}, {Carniani},
  {Marconi}, {Cicone}, {Bourne}, {Costa}, {Fabian}, {Ishibashi}, \&
  {Venturi}}]{Fluetsch19}
{Fluetsch}, A., {Maiolino}, R., {Carniani}, S., {et~al.} 2019, \mnras, 483,
  4586

\bibitem[{{Garc{\'\i}a-Burillo} {et~al.}(2014){Garc{\'\i}a-Burillo}, {Combes},
  {Usero}, {Aalto}, {Krips}, {Viti}, {Alonso-Herrero}, {Hunt}, {Schinnerer}, \&
  {Baker}}]{GarciaBurillo14}
{Garc{\'\i}a-Burillo}, S., {Combes}, F., {Usero}, A., {et~al.} 2014, \aap, 567,
  A125

\bibitem[{{Gaspari} \& {S{\c a}dowski}(2017)}]{Gaspari17}
{Gaspari}, M. \& {S{\c a}dowski}, A. 2017, \apj, 837, 149

\bibitem[{{Hamann} {et~al.}(2018){Hamann}, {Chartas}, {Reeves}, \&
  {Nardini}}]{Hamann18}
{Hamann}, F., {Chartas}, G., {Reeves}, J., \& {Nardini}, E. 2018, \mnras, 476,
  943

\bibitem[{{Herrera-Camus} {et~al.}(2019){Herrera-Camus}, {Tacconi}, {Genzel},
  {F{\"o}rster Schreiber}, {Lutz}, {Bolatto}, {Wuyts}, {Renzini}, {Lilly},
  {Belli}, {{\"U}bler}, {Shimizu}, {Davies}, {Sturm}, {Combes}, {Freundlich},
  {Garc{\'{\i}}a-Burillo}, {Cox}, {Burkert}, {Naab}, {Colina}, {Saintonge},
  {Cooper}, {Feruglio}, \& {Weiss}}]{HerreraCamus19}
{Herrera-Camus}, R., {Tacconi}, L., {Genzel}, R., {et~al.} 2019, \apj, 871, 37

\bibitem[{{Ishibashi} \& {Fabian}(2014)}]{Ishibashi&Fabian14}
{Ishibashi}, W. \& {Fabian}, A.~C. 2014, \mnras, 441, 1474

\bibitem[{{Ishibashi} {et~al.}(2018){Ishibashi}, {Fabian}, \&
  {Maiolino}}]{Ishibashi18}
{Ishibashi}, W., {Fabian}, A.~C., \& {Maiolino}, R. 2018, \mnras, 476, 512

\bibitem[{{Kennicutt}(1998)}]{Kennicutt98}
{Kennicutt}, Jr., R.~C. 1998, \apj, 498, 541

\bibitem[{{King} \& {Pounds}(2015)}]{King&Pounds15}
{King}, A. \& {Pounds}, K. 2015, \araa, 53, 115

\bibitem[{{Longinotti} {et~al.}(2015){Longinotti}, {Krongold}, {Guainazzi},
  {Giroletti}, {Panessa}, {Costantini}, {Santos-Lleo}, \&
  {Rodriguez-Pascual}}]{Longinotti15}
{Longinotti}, A.~L., {Krongold}, Y., {Guainazzi}, M., {et~al.} 2015, \apjl,
  813, L39

\bibitem[{{Longinotti} {et~al.}(2018){Longinotti}, {Vega}, {Krongold},
  {Aretxaga}, {Yun}, {Chavushyan}, {Feruglio}, {G{\'o}mez-Ruiz}, {Monta{\~n}a},
  {Le{\'o}n-Tavares}, {Olgu{\'{\i}}n-Iglesias}, {Giroletti}, {Guainazzi},
  {Kotilainen}, {Panessa}, {Zapata}, {Cruz-Gonzalez}, {Pati{\~n}o-{\'A}lvarez},
  {Rosa-Gonzalez}, {Carrami{\~n}ana}, {Carrasco}, {Costantini}, {Dultzin},
  {Guichard}, {Puerari}, \& {Santos-Lleo}}]{Longinotti18}
{Longinotti}, A.~L., {Vega}, O., {Krongold}, Y., {et~al.} 2018, \apjl, 867, L11

\bibitem[{{Luminari} {et~al.}(2018){Luminari}, {Piconcelli}, {Tombesi},
  {Zappacosta}, {Fiore}, {Piro}, \& {Vagnetti}}]{Luminari18}
{Luminari}, A., {Piconcelli}, E., {Tombesi}, F., {et~al.} 2018, \aap, 619, A149

\bibitem[{{Maiolino} {et~al.}(2012){Maiolino}, {Gallerani}, {Neri}, {Cicone},
  {Ferrara}, {Genzel}, {Lutz}, {Sturm}, {Tacconi}, {Walter}, {Feruglio},
  {Fiore}, \& {Piconcelli}}]{Maiolino12}
{Maiolino}, R., {Gallerani}, S., {Neri}, R., {et~al.} 2012, \mnras, 425, L66

\bibitem[{{McMullin} {et~al.}(2007){McMullin}, {Waters}, {Schiebel}, {Young},
  \& {Golap}}]{McMullin07}
{McMullin}, J.~P., {Waters}, B., {Schiebel}, D., {Young}, W., \& {Golap}, K.
  2007, in Astronomical Society of the Pacific Conference Series, Vol. 376,
  Astronomical Data Analysis Software and Systems XVI, ed. R.~A. {Shaw},
  F.~{Hill}, \& D.~J. {Bell}, 127

\bibitem[{{Moser} {et~al.}(2016){Moser}, {Krips}, {Busch}, {Scharw{\"a}chter},
  {K{\"o}nig}, {Eckart}, {Smaji{\'c}}, {Garc{\'{\i}}a-Marin}, {Valencia-S.},
  {Fischer}, \& {Dierkes}}]{Moser16}
{Moser}, L., {Krips}, M., {Busch}, G., {et~al.} 2016, \aap, 587, A137

\bibitem[{{Nardini} {et~al.}(2015){Nardini}, {Reeves}, {Gofford}, {Harrison},
  {Risaliti}, {Braito}, {Costa}, {Matzeu}, {Walton}, {Behar}, {Boggs},
  {Christensen}, {Craig}, {Hailey}, {Matt}, {Miller}, {O'Brien}, {Stern},
  {Turner}, \& {Ward}}]{Nardini15}
{Nardini}, E., {Reeves}, J.~N., {Gofford}, J., {et~al.} 2015, Science, 347, 860

\bibitem[{{Nardini} \& {Zubovas}(2018)}]{Nardini18}
{Nardini}, E. \& {Zubovas}, K. 2018, \mnras, 478, 2274

\bibitem[{{Pereira-Santaella} {et~al.}(2018){Pereira-Santaella}, {Colina},
  {Garc{\'{\i}}a-Burillo}, {Combes}, {Emonts}, {Aalto}, {Alonso-Herrero},
  {Arribas}, {Henkel}, {Labiano}, {Muller}, {Piqueras L{\'o}pez}, {Rigopoulou},
  \& {van der Werf}}]{Pereira-Santaella18}
{Pereira-Santaella}, M., {Colina}, L., {Garc{\'{\i}}a-Burillo}, S., {et~al.}
  2018, \aap, 616, A171

\bibitem[{{Querejeta} {et~al.}(2017){Querejeta}, {Schinnerer},
  {Garc{\'{\i}}a-Burillo}, {Bigiel}, {Blanc}, {Colombo}, {Hughes}, {Kreckel},
  {Leroy}, {Meidt}, {Meier}, {Pety}, \& {Sliwa}}]{Querejeta17}
{Querejeta}, M., {Schinnerer}, E., {Garc{\'{\i}}a-Burillo}, S., {et~al.} 2017,
  \aap, 599, C1

\bibitem[{{Reeves} {et~al.}(2016){Reeves}, {Braito}, {Nardini}, {Behar},
  {O'Brien}, {Tombesi}, {Turner}, \& {Costa}}]{Reeves16}
{Reeves}, J.~N., {Braito}, V., {Nardini}, E., {et~al.} 2016, \apj, 824, 20

\bibitem[{{Reeves} {et~al.}(2003){Reeves}, {O'Brien}, \& {Ward}}]{Reeves03}
{Reeves}, J.~N., {O'Brien}, P.~T., \& {Ward}, M.~J. 2003, \apjl, 593, L65

\bibitem[{{Rupke} {et~al.}(2005){Rupke}, {Veilleux}, \& {Sanders}}]{Rupke05}
{Rupke}, D.~S., {Veilleux}, S., \& {Sanders}, D.~B. 2005, \apjs, 160, 115

\bibitem[{{Rupke} {et~al.}(2017){Rupke}, {G{\"u}ltekin}, \&
  {Veilleux}}]{Rupke17}
{Rupke}, D.~S.~N., {G{\"u}ltekin}, K., \& {Veilleux}, S. 2017, \apj, 850, 40

\bibitem[{{Rupke} \& {Veilleux}(2013)}]{Rupke&Veilleux13}
{Rupke}, D.~S.~N. \& {Veilleux}, S. 2013, \apj, 768, 75

\bibitem[{{Sakamoto} {et~al.}(2010){Sakamoto}, {Aalto}, {Evans}, {Wiedner}, \&
  {Wilner}}]{Sakamoto10}
{Sakamoto}, K., {Aalto}, S., {Evans}, A.~S., {Wiedner}, M.~C., \& {Wilner},
  D.~J. 2010, \apjl, 725, L228

\bibitem[{{Saturni} {et~al.}(2018){Saturni}, {Bischetti}, {Piconcelli},
  {Bongiorno}, {Cicone}, {Feruglio}, {Fiore}, {Gallerani}, {Giustini},
  {Piranomonte}, {Vietri}, \& {Vignali}}]{Saturni18}
{Saturni}, F.~G., {Bischetti}, M., {Piconcelli}, E., {et~al.} 2018, \aap, 617,
  A118

\bibitem[{{Sijacki} {et~al.}(2007){Sijacki}, {Springel}, {Di Matteo}, \&
  {Hernquist}}]{Sijacki07}
{Sijacki}, D., {Springel}, V., {Di Matteo}, T., \& {Hernquist}, L. 2007,
  \mnras, 380, 877

\bibitem[{{Simpson} {et~al.}(1999){Simpson}, {Ward}, {O'Brien}, \&
  {Reeves}}]{Simpson99}
{Simpson}, C., {Ward}, M., {O'Brien}, P., \& {Reeves}, J. 1999, \mnras, 303,
  L23

\bibitem[{{Solomon} {et~al.}(1997){Solomon}, {Downes}, {Radford}, \&
  {Barrett}}]{Solomon97}
{Solomon}, P.~M., {Downes}, D., {Radford}, S.~J.~E., \& {Barrett}, J.~W. 1997,
  \apj, 478, 144

\bibitem[{{Solomon} \& {Vanden Bout}(2005)}]{Solomon05}
{Solomon}, P.~M. \& {Vanden Bout}, P.~A. 2005, \araa, 43, 677

\bibitem[{{Tacconi} {et~al.}(2013){Tacconi}, {Neri}, {Genzel}, {Combes},
  {Bolatto}, {Cooper}, {Wuyts}, {Bournaud}, {Burkert}, {Comerford}, {Cox},
  {Davis}, {F{\"o}rster Schreiber}, {Garc{\'{\i}}a-Burillo}, {Gracia-Carpio},
  {Lutz}, {Naab}, {Newman}, {Omont}, {Saintonge}, {Shapiro Griffin}, {Shapley},
  {Sternberg}, \& {Weiner}}]{Tacconi13}
{Tacconi}, L.~J., {Neri}, R., {Genzel}, R., {et~al.} 2013, \apj, 768, 74

\bibitem[{{Thompson} {et~al.}(2015){Thompson}, {Fabian}, {Quataert}, \&
  {Murray}}]{Thompson15}
{Thompson}, T.~A., {Fabian}, A.~C., {Quataert}, E., \& {Murray}, N. 2015,
  \mnras, 449, 147

\bibitem[{{Tombesi} {et~al.}(2012){Tombesi}, {Cappi}, {Reeves}, \&
  {Braito}}]{Tombesi12}
{Tombesi}, F., {Cappi}, M., {Reeves}, J.~N., \& {Braito}, V. 2012, \mnras, 422,
  L1

\bibitem[{{Tombesi} {et~al.}(2015){Tombesi}, {Mel{\'e}ndez}, {Veilleux},
  {Reeves}, {Gonz{\'a}lez-Alfonso}, \& {Reynolds}}]{Tombesi15}
{Tombesi}, F., {Mel{\'e}ndez}, M., {Veilleux}, S., {et~al.} 2015, \nat, 519,
  436

\bibitem[{{Torres} {et~al.}(1997){Torres}, {Quast}, {Coziol}, {Jablonski}, {de
  la Reza}, {L{\'e}pine}, \& {Greg{\'o}rio-Hetem}}]{Torres97}
{Torres}, C.~A.~O., {Quast}, G.~R., {Coziol}, R., {et~al.} 1997, \apjl, 488,
  L19

\bibitem[{{Valiante} {et~al.}(2014){Valiante}, {Schneider}, {Salvadori}, \&
  {Gallerani}}]{Valiante14}
{Valiante}, R., {Schneider}, R., {Salvadori}, S., \& {Gallerani}, S. 2014,
  \mnras, 444, 2442

\bibitem[{{Vayner} {et~al.}(2017){Vayner}, {Wright}, {Murray}, {Armus},
  {Larkin}, \& {Mieda}}]{Vayner17}
{Vayner}, A., {Wright}, S.~A., {Murray}, N., {et~al.} 2017, \apj, 851, 126

\bibitem[{{Veilleux} {et~al.}(2017){Veilleux}, {Bolatto}, {Tombesi},
  {Mel{\'e}ndez}, {Sturm}, {Gonz{\'a}lez-Alfonso}, {Fischer}, \&
  {Rupke}}]{Veilleux17}
{Veilleux}, S., {Bolatto}, A., {Tombesi}, F., {et~al.} 2017, \apj, 843, 18

\bibitem[{{Voit} {et~al.}(2015){Voit}, {Donahue}, {O'Shea}, {Bryan}, {Sun}, \&
  {Werner}}]{Voit15}
{Voit}, G.~M., {Donahue}, M., {O'Shea}, B.~W., {et~al.} 2015, \apjl, 803, L21

\bibitem[{{Wang} {et~al.}(2013){Wang}, {Wagg}, {Carilli}, {Walter}, {Lentati},
  {Fan}, {Riechers}, {Bertoldi}, {Narayanan}, {Strauss}, {Cox}, {Omont},
  {Menten}, {Knudsen}, {Neri}, \& {Jiang}}]{Wang13}
{Wang}, R., {Wagg}, J., {Carilli}, C.~L., {et~al.} 2013, \apj, 773, 44

\bibitem[{{Xia} {et~al.}(2012){Xia}, {Gao}, {Hao}, {Tan}, {Mao}, {Omont},
  {Flaquer}, {Leon}, \& {Cox}}]{Xia12}
{Xia}, X.~Y., {Gao}, Y., {Hao}, C.-N., {et~al.} 2012, \apj, 750, 92

\bibitem[{{Yun} {et~al.}(2004){Yun}, {Reddy}, {Scoville}, {Frayer}, {Robson},
  \& {Tilanus}}]{Yun04}
{Yun}, M.~S., {Reddy}, N.~A., {Scoville}, N.~Z., {et~al.} 2004, \apj, 601, 723

\bibitem[{{Zschaechner} {et~al.}(2016){Zschaechner}, {Walter}, {Bolatto},
  {Farina}, {Kruijssen}, {Leroy}, {Meier}, {Ott}, \& {Veilleux}}]{Zchaechner16}
{Zschaechner}, L.~K., {Walter}, F., {Bolatto}, A., {et~al.} 2016, \apj, 832,
  142

\bibitem[{{Zubovas} \& {King}(2012)}]{Zubovas&King12}
{Zubovas}, K. \& {King}, A. 2012, \apjl, 745, L34

\end{thebibliography}

\end{document}